\documentclass[onecolumn]{emulateapj}
\usepackage{amsmath,amssymb,graphicx,natbib,subfigure}

\newcommand{\mh}{{M_\bullet}}
\newcommand{\msun}{{M_\odot}}

\newcommand{\beq}{\begin{equation}}
\newcommand{\eeq}{\end{equation}}
\newcommand{\sbh}{SBH}

\begin{document}

\title{Gravitational Encounters and the Evolution of Galactic Nuclei. I. Method}

\author{David Merritt}
\affil{School of Physics and Astronomy and Center for Computational Relativity
and Gravitation, Rochester Institute of Technology, Rochester, NY 14623}
\email{merritt@astro.rit.edu}

\begin{abstract}
An algorithm is described for evolving the phase-space density of stars 
or compact objects around a massive black hole at the center of a galaxy.
The technique is based on numerical integration of the
Fokker-Planck equation in energy-angular momentum space, $f(E,L,t)$,
and includes, for the first time, diffusion coefficients that describe the effects of both random and 
correlated encounters (``resonant relaxation''), as well as energy loss due to
emission of gravitational waves.
Destruction or loss of stars into the black hole are treated by means of a
detailed boundary-layer analysis.
Performance of the algorithm is illustrated by calculating two-dimensional, 
time-dependent and steady-state distribution functions and their corresponding loss rates.
\end{abstract}

\section{Introduction}

The distribution of stars around a massive black hole is a well-studied problem.
Many solutions are consistent with Jeans's theorem, but after a sufficiently long time,
gravitational encounters between stars will drive their distribution toward a predictable form.
In the case of stars of a single mass, random encounters result in a
steady-state distribution that was first correctly described by Bahcall and Wolf (1976):
the density of stars falls off as $n(r)\propto r^{-7/4}$, and the phase-space density is 
$f(E) \propto |E|^{1/4}$; here $r$ is  distance from the black hole and $E$ is  
the orbital energy per unit mass of a star.
Bahcall and Wolf's derivation assumed that the gravitational potential was dominated
by the black hole, and that the distribution of stars in velocity space was isotropic.
Capture of stars by the hole was allowed to occur only via diffusion in energy;
the steady-state feeding rate was found to be small, in the sense that a small fraction of stars,
within any radius, are consumed by the hole in one, two-body relaxation time at that radius.
For this reason, $f\sim |E|^{1/4}$ is often described as a ``zero flux'' solution
and in fact it can most easily be derived by setting to zero the energy-space flux 
in the evolution equation for the orbital distribution.

\citet{FrankRees1976} argued that feeding of the black hole would be dominated
by diffusion in angular momentum, not energy, and that a fraction of order unity
of stars within radius $r$ would be captured in one relaxation time at $r$.
\citet{LightmanShapiro1977} and \citet{CohnKulsrud1978}  found steady-state solutions
of the equations describing diffusion in both $E$ and $L$; capture
or destruction of stars was assumed to occur when the orbital eccentricity was
large enough, at a given energy, that the orbit intersected the loss sphere,
the region around the hole in which stars are consumed or tidally disrupted. 
The steady-state solutions found by these authors were anisotropic with respect to velocity, 
since the phase space density must be zero near the loss sphere.
However the steady-state energy distribution was found to be quite similar
to that of the scale-free Bahcall-Wolf solution, at least for energies much greater than that of
a circular orbit at the edge of the capture sphere, and the corresponding configuration-space
density was close to $n\sim r^{-7/4}$ outside of this sphere.

These early treatments were made possible by the availability of analytic expressions for the
diffusion coefficients that describe encounter-driven changes in $E$ and $L$.
Such expressions can be straightforwardly derived in the case of random encounters in 
a homogeneous medium \citep[e.g.][]{Rosenbluth1957,CohnKulsrud1978}.
Sufficiently near to a massive black hole -- very roughly, at distances less than $\sim 10^{-1}$ times
the black hole's gravitational influence radius -- the assumption of random encounters breaks down,
since orbits resemble closed Keplerian ellipses for many periods.
In this regime, diffusion in angular momentum is dominated by ``resonant relaxation'' (RR)\citep{RauchTremaine1996} while changes in energy are still controlled by 
random (``non-resonant,'' NR) encounters.
Approximate timescales for changes in $L$ in the RR regime have long been available,
but until recently, usefully accurate expressions for the diffusion coefficients in $L$
were not available.
Happily, that situation has now begun to change \citep{Hamers2014}, 
and so it is feasible to repeat calculations like those of \citet{CohnKulsrud1978} using
expressions for the diffusion coefficients that are valid much closer to the black hole.

This paper, which is the first in a series, presents a numerical algorithm for solving
the Fokker-Planck equation describing the evolution due to gravitational encounters
of $f(E,L)$, the two-dimensional phase-space density of stars orbiting around a massive black hole.
The basic numerical approach is the same as that of \citet{CohnKulsrud1978},
although their algorithm has been generalized to include diffusion coefficients that 
describe both random and correlated encounters.
Cohn \& Kulsrud's treatment is improved upon in other ways as well; notably by the
use of a logarithmic grid in angular momentum, which is necessary to accurately treat 
the behavior of highly eccentric orbits.

As far as we are aware, the time-dependent, $f(E,L,t)$  solutions presented here
are the first ever published.
While Cohn \& Kulsrud's algorithm was capable of calculating time-dependent
solutions, those authors (perhaps because of computer limitations) 
chose to present only steady-state solutions in their 1978 paper.
Time dependence is particularly relevant to galactic nuclei since (energy) relaxation times
are believed to be comparable with galaxy lifetimes, even in nuclei as dense as that
of the Milky Way \citep{Merritt2010}.
This time dependence has routinely been ignored in calculations of the rate of stellar
tidal disruptions \citep[e.g.][]{MagorrianTremaine1998}.

In spite of the improvements, the algorithm presented here still contains some basic limitations.
The mass of the black hole is fixed and its spin is ignored.
Stars are assumed to have a single mass, and the contribution of the stars to the
gravitational potential is ignored.
As a consequence, results are limited in their applicability to a region inside the
black hole's sphere of gravitational influence, 
and the possible influence of spatial asymmetries in the stellar potential on the behavior
of orbits is ignored.
These is no ``source term'' corresponding to star formation and
binary stars are not allowed.
Rotation of the stellar cluster is ignored.
Some of these restrictions will be lifted in later papers from this series.

\section{Assumptions and basic relations}

Stars -- a term used here to refer both to normal stars and to compact objects, e.g. stellar-mass
black holes -- are assumed to have a single mass, $m_\star$.
The stars are assumed to be close enough to the black hole (SBH) that the gravitational 
potential defining their unperturbed orbits is constant in time and given by
\beq
\Phi(r) \equiv -\psi(r) = -\frac{G\mh}{r} 
\eeq
with $\mh$ the \sbh\ mass, also assumed constant.
The contribution to the potential from the stars themselves is ignored, hence
results are only expected to be accurate inside the \sbh\ gravitational influence sphere.

Unperturbed orbits respect the two isolating integrals $E$ (energy) and
$L$ (angular momentum), both defined per unit mass:
\begin{equation}
E = \frac{v^2}{2} +\Phi(r) = \frac{v^2}{2} - \frac{G\mh}{r},\ \  \ \ 
L = |\boldsymbol{r}\times\boldsymbol{v}| .
\end{equation}
Following \citet{CohnKulsrud1978} we define  new variables
(${\cal E}, {\cal R}$) as
\begin{eqnarray}
{\cal E} \equiv -E = -\frac{v^2}{2} +\psi(r), \ \ 
{\cal R} \equiv \frac{L^2}{L_c^2} 
\end{eqnarray}
with $L_c({\cal E})$ the angular momentum of a circular orbit of energy ${\cal E}$:
\beq
L_c^2({\cal E}) = \frac{\left(G\mh\right)^2}{2{\cal E}} .
\eeq
Hence $0\le {\cal R}\le 1$.
${\cal E}$ and ${\cal R}$ are related to the semimajor axis $a$ and eccentricity $e$ 
of the Keplerian orbit via
\beq\label{Equation:semivsE}
a = \frac{G\mh}{2{\cal E}},\ \ \ \ e^2 = 1 - {\cal R} .
\eeq
The orbital (Kepler) period is
\beq
P(a) = \frac{2\pi a^{3/2}}{\sqrt{G\mh}}
 = \frac{\pi}{\sqrt{2}} \frac{G\mh}{{\cal E}^{3/2}}
\eeq
and the gravitational radius of the SBH is
\beq
r_g \equiv \frac{G\mh}{c^2}\approx 4.80\times 10^{-5}
\left(\frac{\mh}{10^6 \msun}\right) \mathrm{mpc} .
\eeq
Spin of the \sbh\ is ignored.

Stars are assumed to be lost -- either captured, or destroyed -- if their
(Newtonian) periapsis distance falls below $r_\mathrm{lc}$, i.e. if
\beq\label{Equation:CaptureCondae}
a(1-e) \le r_\mathrm{lc}.
\eeq
For stars not subject to tidal disruption, e.g. compact objects, $r_\mathrm{lc}\approx 8 r_g$
\citep[][\S 4.6]{DEGN};
otherwise $r_\mathrm{lc}>8r_g$ is the tidal disruption radius.
Defining
\beq\label{Equation:Elc}
{\cal E}_\mathrm{lc} \equiv \frac{G\mh}{2r_\mathrm{lc}} ,
\eeq
 the energy of a circular orbit
at $r=r_\mathrm{lc}$, allows the loss condition to be written as
\begin{eqnarray}\label{Equation:CaptureCond}
{\cal R} &\le& {\cal R}_\mathrm{lc} ({\cal E}) = 
2\frac{\cal E}{{\cal E}_\mathrm{lc}} \left(1-\frac12\frac{\cal E}{{\cal E}_\mathrm{lc}}\right),\ \ 
{\cal E} \le {\cal E}_\mathrm{lc}.
\end{eqnarray}
For ${\cal E}\ll {\cal E}_\mathrm{lc}$, 
equations~(\ref{Equation:Elc}) and (\ref{Equation:CaptureCond}) imply
\beq\label{Equation:CaptureCondApprox}
{\cal R}_\mathrm{lc} \approx 2\frac{\cal E}{\cal E}_\mathrm{lc} \approx 32 \frac{{\cal E}}{c^2} ;
\eeq
the final expression assumes ${\cal E}_\mathrm{lc} = c^2/16$, appropriate for
$r_\mathrm{lc} = 8 r_g$.

The number density of stars in phase space, $f$, is assumed to satisfy
Jeans's theorem at any given time, but the dependence of $f$ on
${\cal E}$ and ${\cal R}$ is allowed to vary with time as gravitational encounters
cause stars to change their orbits:
\begin{eqnarray}
f&=&f\left({\cal E}, {\cal R}, t\right) .
\end{eqnarray}
The configuration-space density is given in terms of  $f$ as
\begin{subequations}\label{Equation:nofr}
\begin{eqnarray}
n(r) &=& \frac{2\pi}{r^2} \int_0^{\psi(r)} L_c^2({\cal E})\; d{\cal E}
\int_0^{{\cal R}_\mathrm{max}} 
\frac{f({\cal E}, {\cal R}) d{\cal R}}
{\sqrt{2\left[\psi(r)-{\cal E}\right] - {\cal R}L_c^2({\cal E},t)/r^2}} \\
&=& \sqrt{2}\pi \frac{G\mh}{r} \int_0^{G\mh/r} \frac{d{\cal E}}{\sqrt{\cal E}}
\int_0^{{\cal R}_\mathrm{max}} 
\frac{f({\cal E}, {\cal R},t) d{\cal R}}
{\sqrt{{\cal R}_\mathrm{max} - {\cal R}}} 
\end{eqnarray}
\end{subequations}
where
\beq
{\cal R}_\mathrm{max} ({\cal E},r) = \frac{2r^2\left[\psi(r)-{\cal E}\right]}{L_c^2({\cal E})}
= 4 \left(\frac{r {\cal E}}{G\mh}\right)\left(1-\frac{r {\cal E}}{G\mh}\right) 
\eeq
with similar expressions for the velocity dispersions $\sigma_r, \sigma_t$ in the
radial and transverse directions.

The time dependence of $f$ is assumed to be described by the 
orbit-averaged Fokker-Planck equation:
\begin{eqnarray}\label{Equation:FPOAEL}
\frac{\partial N}{\partial t} =-\frac{\partial}{\partial {\cal E}}\left(N\langle\Delta {\cal E}\rangle_t\right) + \frac12\frac{\partial^2}{\partial {\cal E}^2}\left(N\langle(\Delta {\cal E})^2\rangle_t\right) - \frac{\partial}{\partial {\cal R}}\left(N\langle\Delta {\cal R}\rangle_t\right)
+ \frac12\frac{\partial^2}{\partial {\cal R}^2}\left(N\langle(\Delta {\cal R})^2\rangle_t\right)
+ \frac{\partial^2}{\partial {\cal E}\partial {\cal R}}\left(N\langle\Delta {\cal E}\Delta {\cal R}\rangle_t\right)\nonumber \\
\end{eqnarray}
where 
\begin{eqnarray}
N({\cal E}, {\cal R},t) &=& 4\pi^2 L_c^2({\cal E}) P({\cal E}, {\cal R}) f({\cal E}, {\cal R}, t) 
= \sqrt{2}\pi^3\left(G\mh\right)^3 {\cal E}^{-5/2} f({\cal E}, {\cal R}, t) 
\equiv {\cal J}({\cal E}, {\cal R}) f({\cal E}, {\cal R}, t)
\end{eqnarray} 
is the distribution of orbital integrals.
The quantities in $\langle \rangle$ in equation (\ref{Equation:FPOAEL}) are diffusion coefficients;
the subscript ``$t$'' indicates a time average over the unperturbed orbit.
The diffusion coefficients are functions of ${\cal E}$, ${\cal R}$ and of the stellar
distribution itself, i.e. of $f$; their functional forms are discussed in more detail below.

Equation~(\ref{Equation:FPOAEL}) can be written in flux-conservation form as
\citep{CohnKulsrud1978}
\begin{eqnarray}
{\cal J}\frac{\partial f}{\partial t} &=& 
-\frac{\partial}{\partial{\cal E}}\left({\cal J}\phi_{\cal E}\right) 
- {\cal J}\frac{\partial}{\partial {\cal R}}\phi_{\cal R} \nonumber \\
-\phi_{\cal E} &=& D_{\cal E\cal E}\frac{\partial f}{\partial {\cal E}} + 
D_{\cal E\cal R} \frac{\partial f}{\partial {\cal R}} +
D_{\cal E} f,\nonumber \\
-\phi_{\cal R} &=& D_{\cal R\cal E}\frac{\partial f}{\partial {\cal E}} + 
D_{\cal R\cal R} \frac{\partial f}{\partial {\cal R}} + D_{\cal R} f
\label{Equation:FPFluxConserve}
\end{eqnarray}
with ``flux coefficients''
\begin{eqnarray}\label{Equation:DefineFluxCoefs}
D_{\cal E} &=& -\langle\Delta{\cal E}\rangle_t 
- \frac{5}{4{\cal E}}\langle\left(\Delta{\cal E}\right)^2\rangle_t 
+ \frac12\frac{\partial}{\partial{\cal E}}\langle\left(\Delta{\cal E}\right)^2\rangle_t 
+ \frac12\frac{\partial}{\partial{\cal R}}\langle\Delta{\cal E}\Delta {\cal R}\rangle_t \;,
\nonumber \\
D_{\cal R} &=& -\langle\Delta{\cal R}\rangle_t 
- \frac{5}{4{\cal E}}\langle\Delta{\cal E}\Delta{\cal R}\rangle_t 
+ \frac12\frac{\partial}{\partial{\cal E}}\langle\Delta{\cal E}\Delta{\cal R}\rangle_t 
+ \frac12\frac{\partial}{\partial{\cal R}}\langle\left(\Delta {\cal R}\right)^2\rangle_t \;,
\nonumber \\
D_{\cal E\cal E} &=& \frac12 \langle\left(\Delta{\cal E}\right)^2\rangle_t \;,
D_{\cal E\cal R} = D_{\cal R\cal E} = \frac12\langle\Delta{\cal E}\Delta{\cal R}\rangle_t\; ,
D_{\cal R\cal R} = \frac12\langle\left(\Delta{\cal R}\right)^2\rangle_t \; .
\end{eqnarray}

The rate of loss of stars past the loss-cone boundary:
\begin{equation}
{\cal R} = {\cal R}_\mathrm{lc}({\cal E}), \ \ 
{\cal E}_\mathrm{min} \le {\cal E} \le {\cal E}_\mathrm{max}
\end{equation}
is given in terms of the fluxes as
\begin{subequations}\label{Equation:Ndot}
\begin{eqnarray}
\dot N &=& \int_{{\cal E}_\mathrm{min}}^{{\cal E}_\mathrm{max}} d{\cal E} \int_{{\cal R}_\mathrm{lc}({\cal E})}^1  {\cal J} \frac{\partial f}{\partial t}
d{\cal R} 
= -\int_{{\cal E}_\mathrm{min}}^{{\cal E}_\mathrm{max}} d{\cal E}  \int_{{\cal R}_\mathrm{lc}({\cal E})}^1  \left[\frac{\partial}{\partial {\cal E}}
\left({\cal J} \phi_{\cal E}\right) + {\cal J}\frac{\partial}{\partial {\cal R}}\phi_{\cal R}\right] 
\label{Equation:Ndota} \\
&=& \int_{{\cal E}_\mathrm{min}}^{{\cal E}_\mathrm{max}} {\cal J}({\cal E})\;
 \phi_{\cal R}\left[{\cal E}, {\cal R}_\mathrm{lc}\left({\cal E}\right)\right] d{\cal E} -
\int_{{\cal R}_\mathrm{min}} ^1 {\cal J}\left({\cal E}_\mathrm{lc}\right) \; \phi_{\cal E}\left[{\cal E}_\mathrm{lc}\left({\cal R}\right), {\cal R}\right] d{\cal R}.
\label{Equation:Ndotb}
\end{eqnarray}
\end{subequations}
In the final expression, ${\cal R}_\mathrm{min} \equiv {\cal R}_\mathrm{lc}({\cal E}_\mathrm{min})$
and ${\cal E}_\mathrm{lc}$ is the value of ${\cal E}$ satisfying ${\cal R} = {\cal R}_\mathrm{lc}({\cal E})$.
We expect the first of the two terms in (\ref{Equation:Ndotb}) (flux due to diffusion in $L$)
to dominate the loss rate.

A quantity that plays an important role in the angular momentum diffusion of orbits
near the loss-cone boundary is
\begin{equation}\label{Equation:calDofE}
{\cal D}({\cal E}) \equiv
\lim_{{\cal R} \rightarrow 0} \frac{\langle\left(\Delta{\cal R}\right)^2\rangle_t}{2{\cal R}}
= \lim_{{\cal R} \rightarrow 0} \frac{D_{\cal R R} ({\cal E, R})}{\cal R}.
\end{equation}
In this limit, and ignoring diffusion in ${\cal E}$, the Fokker-Planck equation becomes
\begin{equation}
\frac{\partial N}{\partial \left({\cal D}t\right)} \approx 
\frac{\partial}{\partial {\cal R}}\left({\cal R}\frac{\partial N}{\partial {\cal R}}\right)
\end{equation}
showing that ${\cal D}^{-1}$ is effectively an orbit-averaged, angular momentum relaxation time
at energy ${\cal E}$.
Another quantity that can be expressed in terms of ${\cal D}$ is
\begin{eqnarray}\label{Equation:Defineqlc}
q_\mathrm{lc}({\cal E}) \equiv  \frac{P({\cal E}) {\cal D}({\cal E})}{{\cal R}_\mathrm{lc}({\cal E})}.
\end{eqnarray}
where
$q_\mathrm{lc}\gtrsim 1$ defines the full-loss-cone regime.
Since the limiting value of ${\cal D}$ as ${\cal R}\rightarrow 0$ may not be well defined
it is more useful to define ${\cal D}$ as
\begin{equation}\label{Equation:calDofElc}
{\cal D}({\cal E}) \equiv
\frac{\langle\left(\Delta{\cal R}\right)^2\rangle_t}{2{\cal R}}\bigg|_{{\cal R}={\cal R}_\mathrm{lc}}
= \frac{D_{\cal R R} ({\cal E}, {\cal R}_\mathrm{lc})}{{\cal R}_\mathrm{lc}}.
\end{equation}
\section{Diffusion coefficients}
\bigskip
\subsection{Cohn-Kulsrud diffusion coefficients}
Orbit-averaged diffusion coefficients were derived by \citet{CohnKulsrud1978} for stars moving in a Kepler potential.
Their derivation was based on the theory of random gravitational encounters as developed by 
Chandrasekhar, H\'enon, Spitzer and others.
Cohn \& Kulsrud wrote their orbit-averaged diffusion coefficients as
\begin{eqnarray}
\langle\Delta{\cal E}\rangle_t &=& -F_0 + F_1\; ,\ \ \ \ 
\langle\left(\Delta{\cal E}\right)^2\rangle_t = \frac43{\cal E} F_0 + \frac43 {\cal E} F_4, 
\ \ \ \ 
\langle\Delta{\cal E}\Delta{\cal R}\rangle_t = \frac{4\cal R}{3} F_4 - \frac{4\cal R}{3} F_5 ,
\nonumber \\
\langle\Delta{\cal R}\rangle_t &=& \frac{5}{3{\cal E}}\left(1-2{\cal R}\right) F_0 + 
\frac{\cal R}{\cal E} F_1 - \frac52\frac{\cal R}{\cal E} F_2 + \frac{4}{\cal E} F_3 -
\frac43\frac{\cal R}{\cal E} F_5 + \frac{\cal R}{2\cal E} F_6 - \frac{4}{3\cal E} F_7, 
\nonumber \\
\langle\left(\Delta{\cal R}\right)^2\rangle_t &=& 
\frac{10}{3}\frac{\cal R}{\cal E}\left(1-{\cal R}\right) F_0
- 2\frac{{\cal R}^2}{\cal E} F_2 + 8 \frac{\cal R}{\cal E} F_3 +
\frac43 \frac{{\cal R}^2}{\cal E} F_4 - \frac83\frac{{\cal R}^2}{\cal E} F_5
+ 2\frac{{\cal R}^2}{\cal E} F_6 - \frac83\frac{\cal R}{\cal E} F_7 
\label{Equation:NRDiffCoefs}
\end{eqnarray}
where the functions $F_i$, $i=0\ldots 7$ are expressed as integrals depending on $f$.
We follow those authors and assume that the $F_i$
depend only on the angular-momentum average of $f$, or
\beq
\overline{f}({\cal E},t) \equiv \int_0^1 f({\cal E}, {\cal R},t)\; d{\cal R} .
\eeq
With this simplification, Cohn \& Kulsrud show that the $F_i$ are
\begin{subequations}\label{Equation:Fi}
\begin{eqnarray}\label{Equation:Fi0}
F_0\left({\cal E}\right) &=& 4\pi\Gamma\int_0^{\cal E} \overline{f}\left({\cal E}^\prime\right) d{\cal E}^\prime \; , \\
\label{Equation:Fii}
F_i\left({\cal E}, {\cal R}\right) &=& 4\pi\Gamma\int_{\cal E}^{{\cal E}/x_-} 
\overline{f}\left({\cal E}^\prime\right) C_i\left(\frac{{\cal E}^\prime}{\cal E}, {\cal R}\right)
d{\cal E}^\prime \; , 
\end{eqnarray}
\end{subequations}
where $\Gamma \equiv 4\pi G^2 m_\star^2\ln\Lambda$, $\ln\Lambda$ is
the Coulomb logarithm, and
$x_\pm\equiv \frac12\left[1\pm\left(1-{\cal R}\right)^{1/2}\right]$.
The functions $C_i\left(s={\cal E}^\prime/{\cal E}, {\cal R}\right)$ 
 have the forms
\beq
C_i = \frac{2}{\pi} \int dx\; Q^{-1/2}\; \frac{x^l\left(1-sx\right)^{m/2}}{\left(1-x\right)^{n/2}} \nonumber
\eeq
with $(l,m,n)$ integers and $Q\equiv (x_+-x)(x-x_-)$.
Appendix A gives explicit expressions for the $C_i$ and describes how they 
were computed numerically.

Henceforth the diffusion coefficients (\ref{Equation:NRDiffCoefs})
will be referred to as the Cohn-Kulsrud (CK) diffusion coefficients
and given the subscript ``CK." 
The subscript $t$, for orbit-averaging, is understood in everything that follows and will be omitted
henceforth.

The quantity ${\cal D}({\cal E}) \equiv [\langle\left(\Delta{\cal R}\right)^2\rangle_t/(2{\cal R})]_{{\cal R}={\cal R}_\mathrm{lc}}$
defined in equation (\ref{Equation:calDofElc}), which is effectively an orbit-averaged,
angular momentum relaxation time, can be written in terms of the $F_i$ as
\begin{eqnarray}\label{Equation:calDofENR}
{\cal D}({\cal E}) = \frac{1}{3\cal E} \left[5F_0\left({\cal E}\right) + 12 F_3\left({\cal E},{\cal R}_\mathrm{lc}\right) - 4F_7\left({\cal E},{\cal R}_\mathrm{lc}\right)\right] 
\end{eqnarray}
and the quantity $q_\mathrm{lc}$ defined in equation (\ref{Equation:Defineqlc}) is
\begin{eqnarray}\label{Equation:qlcNR}
q_\mathrm{lc}({\cal E}) = \frac{\pi}{3 \sqrt{2}} \frac{G\mh}{{\cal E}^{5/2}}
\frac{1}{{\cal R}_\mathrm{lc}({\cal E})}
\left[5F_0\left({\cal E}\right) + 12 F_3\left({\cal E},{\cal R}_\mathrm{lc}\right) - 4F_7\left({\cal E},{\cal R}_\mathrm{lc}\right)\right] .
\end{eqnarray}

\bigskip
\subsection{Resonant relaxation}
\label{Section:RR}
The theory of random gravitational encounters that is the basis for the Cohn-Kulsrud
diffusion coefficients fails to adequately describe the evolution of 
orbits sufficiently near to a \sbh, where motion is close to Keplerian and where
orbits maintain their orientations for many periods \citep{RauchTremaine1996}.
In this regime, it is common to assume that random gravitational encounters
are still active at changing orbital $E$ and $L$, according to the diffusion coefficients
defined above, but that torques due to the
nearly-fixed Keplerian orbits are also effective at changing $L$, 
sometimes on time scales much shorter than the 
relaxation time defined in terms of random encounters \citep{HopmanAlexander2006,Eilon2009}.

We make the same assumptions here, and write
the diffusion coefficients that appear in the Fokker-Planck equation as 
(orbit-averaging understood)
\begin{eqnarray}\label{Equation:CombinedDiffCoefs}
\langle\Delta{\cal E}\rangle &=&  \langle\Delta{\cal E}\rangle_\mathrm{CK},\ \ \ \ 
\langle\left(\Delta{\cal E}\right)^2\rangle = \langle\left(\Delta{\cal E}\right)^2\rangle_\mathrm{CK}, 
\ \ \ \ 
\langle\Delta{\cal E}\Delta{\cal R}\rangle =  
\langle\Delta{\cal E}\Delta{\cal R}\rangle_\mathrm{CK} ,
\nonumber \\
\langle\Delta{\cal R}\rangle &=& \langle\Delta{\cal R}\rangle_\mathrm{CK}
+ \langle\Delta{\cal R}\rangle_\mathrm{RR}, \ \ \ \ 
\langle\left(\Delta{\cal R}\right)^2\rangle =
\langle\left(\Delta{\cal R}\right)^2\rangle_\mathrm{CK} + 
 \langle\left(\Delta{\cal R}\right)^2\rangle_\mathrm{RR} \;.
\end{eqnarray}
It is understood that the resonant diffusion coefficients describe changes in $L$ in the 
``incoherent'' (as opposed to the ``coherent'') regime, 
i.e., on time scales long compared with the coherence time (defined below).

No very complete theory of incoherent resonant relaxation exists.
While the approximate dependence of the diffusion rate on energy (i.e. distance from the \sbh)
is not difficult to derive, until recently, little was known about the angular-momentum dependence
of the first- and second-order $L$-diffusion coefficients.
We  base what follows on the numerical treatment of \citet{Hamers2014}, who used
an algorithm called {\tt TPI} (``test-particle integrator'') to infer values of the
angular momentum diffusion coefficients for test stars orbiting in nuclei with 
$n(r)\propto r^{-2}$ and $r^{-1}$.

Those authors expressed the diffusion coefficients in terms of the angular momentum variable
\beq
\ell \equiv \frac{L}{L_c(E)} = {\cal R}^{1/2} = \sqrt{1-e^2} \;.
\eeq
A straightforward transformation yields the relations between diffusion coefficients
in $\ell$ and in ${\cal R}$:
\begin{eqnarray}\label{Equation:elltoR}
\langle\Delta{\cal R}\rangle&=& 2\ell\langle\Delta\ell\rangle
+ \langle\left(\Delta\ell\right)^2\rangle ,\ \ \ \ 
\langle\left(\Delta{\cal R}\right)^2\rangle = 4\ell^2  \langle\left(\Delta\ell\right)^2\rangle
\end{eqnarray}
or
\begin{eqnarray}\label{Equation:Rtoell}
\langle\Delta\ell\rangle &=& \frac{1}{2\sqrt{\cal R}} \langle\Delta{\cal R}\rangle -
\frac{1}{8{\cal R}^{3/2}} \langle(\Delta{\cal R})^2\rangle , \ \ \ \ 
\langle(\Delta\ell)^2\rangle = \frac{1}{4{\cal R}} \langle(\Delta{\cal R})^2\rangle \;.
\end{eqnarray}

Hamers et al. (2014) proposed the following forms for the 
the first- and second-order diffusion coefficients:
\begin{subequations}\label{Equation:RRDiffCoef}
\begin{eqnarray}\label{Equation:RRDiffCoefa}
\langle\Delta \ell\rangle_\mathrm{RR} &=& C_1\; A({\cal E})\; g(\ell) \ \ \ \ 
 \langle\left(\Delta\ell\right)^2\rangle_\mathrm{RR} =  C_2\; A({\cal E})\; h(\ell) \\
A(a) &=&  \alpha_s^2\left[\frac{M_\star(a)}{\mh}\right]^2 \frac{1}{N(a)} 
\frac{t_\mathrm{coh}(a)}{P(a)^2} \\
g(\ell) &=& \frac{1}{\ell}\left(1-\frac{\ell^2}{\ell_c^2}\right),
\ \ \ \ h(\ell) = 1-\ell^2 .
\label{Equation:RRDiffCoef2}
\end{eqnarray}
\end{subequations}
In these expressions, $a$ is related to ${\cal E}$ via equation (\ref{Equation:semivsE}).
The value of $\alpha_s$ was determined numerically to be $\sim 1.6$,
and $\ell_c$ was estimated to be $\sim 0.7$, weakly dependent on $a$.
Hamers et al. suggested $C_1\approx C_2\approx 1$.
The quantity $t_\mathrm{coh}(a)$ in equations (\ref{Equation:RRDiffCoef}) is the ``coherence time,'' defined as the
typical time, for stars of semimajor axis $a$, to precess by an angle $\pi$.
Hamers et al. assumed\footnote{Some authors, e.g. \citet{HopmanAlexander2006} and
\citet{Madigan2011},  define the coherence time in terms of a {\it difference} between 
the two precession rates, implying an infinite coherence time at some radius.
The reason why this is incorrect is discussed in \citet{DEGN}, \S5.6.1.}
\begin{eqnarray}\label{Equation:Definetcoh}
t_\mathrm{coh}^{-1} \equiv t_\mathrm{coh,M}^{-1} + t_\mathrm{coh,S}^{-1}
\end{eqnarray}
where
\begin{equation}\label{Equation:Deftcoh}
t_\mathrm{coh,M}(a) = \frac{\mh}{N(a)m_\star}P(a)\;, \ \ \ \ 
t_\mathrm{coh,S}(a) = \frac{1}{12} \frac{a}{r_g} P(a).
\end{equation}
The latter are averages over eccentricity of the mass- and Schwarzschild apsidal precession
times, respectively, assuming a ``thermal'' eccentricity distribution,
$N(e;a)de \propto e de$ \citep[][\S5.6.1.1]{DEGN}.
The Schwarzschild coherence time is independent of the mass distribution and is given by
\begin{eqnarray}\label{Equation:DeftcohN}
t_\mathrm{coh,S}(a) = \frac{\pi}{6} \frac{c^2 a^{5/2}}{\left(G\mh\right)^{3/2}} 
\approx  5.1\times 10^3 \left(\frac{\mh}{10^6\msun}\right)^{-3/2}
\left(\frac{a}{10^{-3} \mathrm{pc}}\right)^{5/2} \mathrm{yr}.
\end{eqnarray}
The mass coherence time depends on the stellar distribution.
These two coherence times are equal when
\beq
a N(a) = 12 \frac{\mh}{m_\star} r_g 
\eeq

In the approximate theory that motivated the expression (\ref{Equation:RRDiffCoef}) for $A(a)$, 
the quantity $N(a)$ could mean either ``number of stars with instantaneous radii less than $a$'' or 
``number of stars with semimajor axes less than $a$''.
In fact, these two functions are quite similar, at least in nuclei with steeply-rising density near
the \sbh\ (Appendix B).
In his numerical experiments, A. Hamers (private communication) was not able to determine which
definition of $N(a)$ provided the better fit to the data.
In what follows, we adopt the first definition, which is simpler to implement in the code: 
\beq
N(a) = 4\pi \int_0^a n(r) r^2 dr 
\eeq
and $M_\star(a) = m_\star N(a)$.

The $a-$ dependence of the diffusion coefficients in equations 
(\ref{Equation:RRDiffCoef}) is based on theoretical arguments, but the $\ell-$ dependence
is essentially ad hoc.
 While the diffusion coefficients derived numerically by Hamers et al. were
consistent with the  $\ell$ dependence of equations (\ref{Equation:RRDiffCoef}),
the functional forms themselves were not strongly constrained.
We now consider those functional forms in more detail.

The variables $\ell$ or ${\cal R}$ are bounded, and therefore the flux in ${\cal R}$:
\beq\label{Equation:FluxR}
\phi_{\cal R} = -D_{\cal R}f - D_{\cal RR}\frac{\partial f}{\partial {\cal R}} 
\eeq
must go to zero as ${\cal R} \rightarrow 0$ or ${\cal R}\rightarrow 1$.
Furthermore this must be true for any $f({\cal R})$ in equation (\ref{Equation:FluxR}).
Hence we require 
\beq\label{Equation:DRzero}
D_{\cal R} \rightarrow 0,\ \ \ \ D_{\cal RR}\rightarrow 0
\eeq
as ${\cal R}\rightarrow \{0,1\}$.
According to equations (\ref{Equation:DefineFluxCoefs}) and (\ref{Equation:elltoR}),
\begin{subequations}\label{Equation:DRDRR}
\begin{eqnarray}
D_{\cal R} &=& -\langle\Delta{\cal R}\rangle + \frac12\frac{\partial}{\partial {\cal R}}\langle(\Delta{\cal R})^2\rangle = -2\ell\langle\Delta\ell\rangle + \langle(\Delta\ell)^2\rangle +
\ell\frac{\partial}{\partial\ell}\langle(\Delta\ell)^2\rangle \\
D_{\cal RR} &=& \frac12\langle(\Delta{\cal R})^2\rangle  = 2\ell^2\langle(\Delta\ell)^2\rangle .
\end{eqnarray}
\end{subequations}
Equations~ (\ref{Equation:DRzero}) - (\ref{Equation:DRDRR}) imply
\begin{subequations}\label{Equation:DRCond}
\begin{eqnarray}
\langle\Delta{\cal R}\rangle &\rightarrow& \frac12\frac{\partial}{\partial {\cal R}}\langle(\Delta{\cal R})^2\rangle ,\ \ \ \ 
\langle(\Delta{\cal R})^2\rangle \rightarrow 0 \\
\ell\langle\Delta\ell\rangle &\rightarrow&
\frac12\frac{\partial}{\partial\ell}\left[\ell\langle(\Delta\ell)^2\rangle\right],\ \ \ \ 
\ell^2\langle(\Delta\ell)^2\rangle \rightarrow 0
\end{eqnarray}
\end{subequations}
as $\{{\cal R}, \ell\}\rightarrow$ 0 or 1.

The expression for $\langle\left(\Delta\ell\right)^2\rangle$ 
in equation (\ref{Equation:RRDiffCoef}) satisfies these conditions.

In the case of the first-order coefficient for $\ell$, 
the condition (\ref{Equation:DRCond}b) becomes
\beq
C_1 \ell g(\ell) = \frac{C_2}{2} \frac{\partial}{\partial \ell} \left[\ell h(\ell)\right],
\ \ \ \ \ell\rightarrow \{0,1\}
\eeq
i.e.
\beq
C_1\left(1-\frac{\ell^2}{\ell_c^2}\right) = \frac{C_2}{2} \left(1 - 3\ell^2\right),
\ \ \ \ \ell\rightarrow \{0,1\}.
\eeq
Applying this respectively at $\ell=\{0,1\}$ yields
\begin{subequations}\label{Equation:Result1}
\begin{eqnarray}
C_1 &=& \frac{C_2}{2} \ \ \ \ \ (\ell=0), \label{Equation:Result1a} \\
C_1\left(1-\ell_c^2\right) &=& C_2 \ell_c^2\ \ \ \ (\ell=1) \label{Equation:Result1b}
\end{eqnarray}
\end{subequations}
and therefore
\beq\label{Equation:ellcone}
\ell_c = \frac{1}{\sqrt{3}} \approx 0.577 \ \ \ \ (C_2=2C_1) .
\eeq
Interestingly, the first-order flux
coefficient implied by these functional forms:
\begin{eqnarray}
D_{\cal R}(\ell) &=& A(E)\left[-2C_1\left(1-\frac{\ell^2}{\ell_c^2}\right) + C_2\left(1-\ell^2\right) + \ell C_2\left(-2\ell\right)\right] \nonumber \\
&=& A(E)\left[\left(C_2-2C_1\right)\ell_c^2 + \left(2C_1-3C_2\ell_c^2\right)\ell^2\right]
\end{eqnarray}
is identically zero if we choose $\left\{C_2=2C_1,\;\ell_c^2=1/3\right\}$.
Hence, imposing $D_{\cal R}=0$ at the boundaries implies $D_{\cal R}=0$
everywhere -- though not necessarily zero {\it flux} everywhere.

Based on their numerical experiments, 
\citet{Hamers2014} suggested a different choice of parameters:
$C_1=C_2$ and $\ell_c\approx 0.7$.
Indeed, setting $C_1=C_2$ in equation (\ref{Equation:Result1b}) yields
\beq\label{Equation:ellctwo}
\ell_c = \frac{1}{\sqrt{2}} \approx 0.707 \ \ \ \ (C_2=C_1) .
\eeq
The implied flux coefficient is
\beq
D_{\cal R} = -\frac{C_1}{2}\left(1-\ell^2\right) ,
\eeq
zero at $\ell=1$ but not at $\ell=0$.

In face of these issues, we returned to the numerical results of Hamers et al. 
and considered more general functional forms for the diffusion coefficients.

The ${\cal R}$-diffusion coefficients implied by equations (\ref{Equation:RRDiffCoef}) have the forms
\begin{subequations}\label{Equation:RDiffusion1}
\begin{eqnarray}
\langle \Delta{\cal R}\rangle &=& A(E) \left[\left(2C_1+C_2\right) - \left(C_2+2C_1/\ell_c^2\right){\cal R}\right], \\
\langle\left(\Delta {\cal R}\right)^2\rangle &=& A(E) \times 4C_2 {\cal R}\left(1-{\cal R}\right),
\end{eqnarray}
\end{subequations}
linear in the case of the first-order coefficient and quadratic in the case of the second-order
coefficient.
Consider the more general functional forms
\begin{subequations}\label{Equation:MoreGeneral}
\begin{eqnarray}
\langle \Delta{\cal R}\rangle &=& A(E) \left(a + b{\cal R} + c{\cal R}^2\right), \\
\langle\left(\Delta {\cal R}\right)^2\rangle &=& A(E) \left(d + e{\cal R} + f{\cal R}^2\right)
\end{eqnarray}
\end{subequations}
which adds an extra parameter, for a total of six.
We now require that these parameters be chosen so as to satisfy all the following  conditions:
\begin{enumerate}
\item $\langle\left(\Delta{\cal R}\right)^2\rangle \ge 0 $
\item $\langle\left(\Delta{\cal R}\right)^2\rangle = 0 $ for ${\cal R}=\{0,1\}$
\item $\langle\Delta{\cal R}\rangle = \frac12\frac{\partial}{\partial {\cal R}}\langle(\Delta{\cal R})^2\rangle$ for ${\cal R}=\{0,1\}$
\end{enumerate}
It is easy to show that these conditions leave only two independent parameters,
allowing the diffusion coefficients (\ref{Equation:MoreGeneral}) to be written as
\begin{subequations}\label{Equation:AnalyticNew}
\begin{eqnarray}
\langle \Delta{\cal R}\rangle &=& A(E) \times C \left[1 - \eta{\cal R} + \left(\eta-2\right){\cal R}^2\right], \\
\langle\left(\Delta {\cal R}\right)^2\rangle &=& A(E)\times 2C\times {\cal R}\left(1-{\cal R}\right) .
\end{eqnarray}
\end{subequations}
The parameter $C$ is a normalization; 
based on comparison with equation~(\ref{Equation:RDiffusion1}),
we expect $C=2C_1+C_2\approx 2$.
The parameter $\eta$ can have any value;
it determines the value of ${\cal R}$ at which $\langle\Delta {\cal R}\rangle=0$,
i.e. the value of $x\equiv {\cal R}_0$ that solves
\beq
\left(\eta-2\right)x^2 - \eta x + 1 = 0
\eeq
(there is only one root in the range $x\subset [0,1]$).
$\eta=\infty$ corresponds to ${\cal R}_0=0$ and
$\eta=-\infty$ to ${\cal R}_0=1$.

A. Hamers kindly provided data files with the numerically-determined diffusion coefficients,
as presented in the Hamers et al. 2014 paper.
The best-fitting values of $\{C,\eta\}$ were determined from these data,
at each semimajor axis bin, as follows:
\begin{enumerate}
\item Given the numerically-computed diffusion coefficients in $\ell$, the
1st- and 2nd-order diffusion coefficients in ${\cal R}$ were computed at each of the data points ${\cal R}_i$ using equations (\ref{Equation:elltoR}).
To simplify the notation, we call these $D_1({\cal R}_i)$
and $D_2({\cal R}_i)$ respectively.
\item Data were excluded if they lay outside the range $\ell_1\le \ell \le \ell_2$,
with $\ell_2=0.95$ and $\ell_1$ equal to 1.1 times the predicted location of the 
``Schwarzschild barrier'' (as defined below).
\item The quantity
$$
\sum_\mathrm{data} \left\{ D_1({\cal R}_i) - C\left[1-\eta{\cal R}_i + \left(\eta-2\right){\cal R}_1^2\right]\right\}^2 + 
\left[D_2({\cal R}_i) - 2C{\cal R}_i\left(1-{\cal R}_i\right)\right]^2
$$
was minimized on a grid in $\{C,\eta\}$.
\end{enumerate}
\begin{figure}[h!]
\begin{center}
  \includegraphics[angle=0.,width=4.0in]{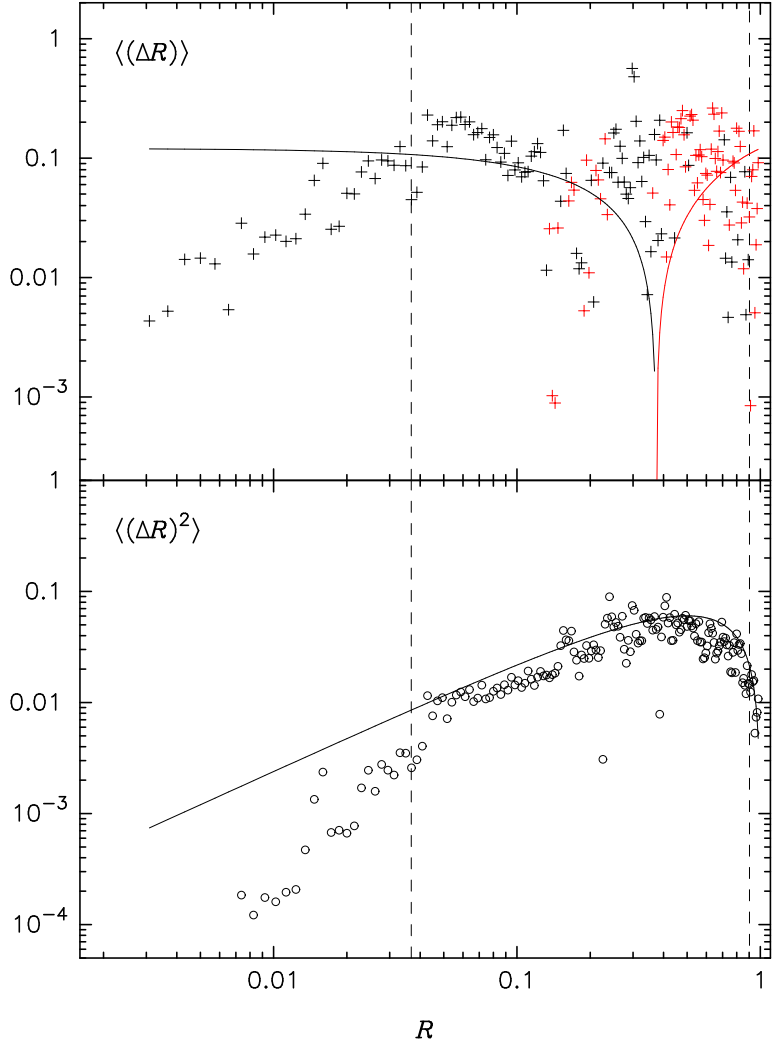}
  \caption{Fits to the diffusion coefficient data from \citet{Hamers2014}
for the radial bin $\langle a \rangle = 19.9$ mpc.
In the upper panel, red symbols are $-\langle\Delta{\cal R}\rangle$.
Dashed vertical lines delineate the region in which data were fit to the 
analytic forms (\ref{Equation:AnalyticNew}); the latter are shown as the solid curves.
}
\label{Figure:dfit}
\end{center}
\end{figure}

Figure \ref{Figure:dfit} shows the fits to the data in the bin $\langle a\rangle = 19.9$ mpc.
\begin{figure}[h!]
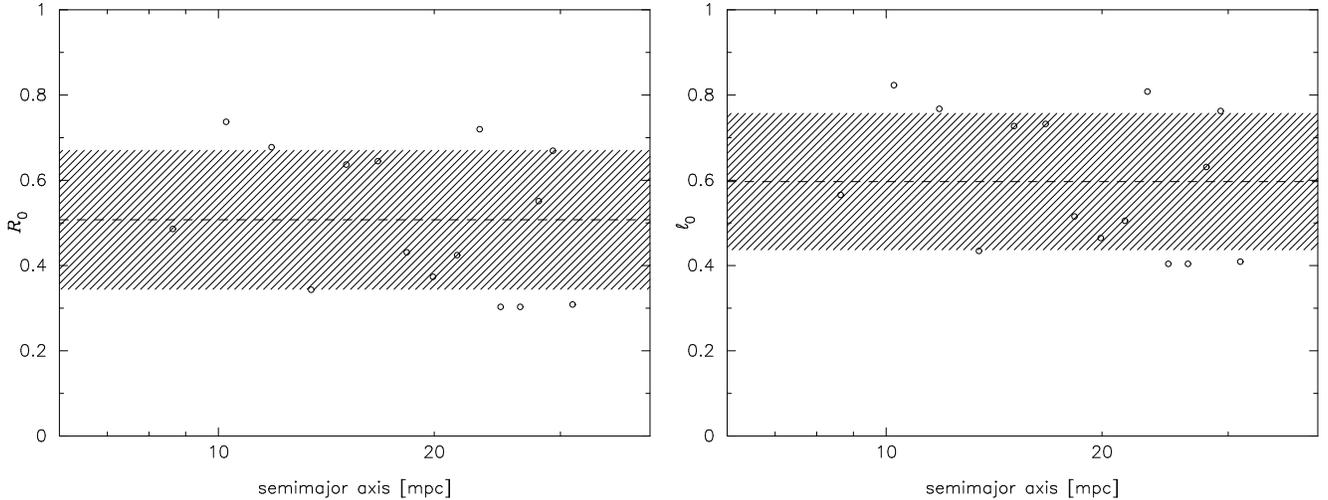

\centering
\mbox{\subfigure{\includegraphics[width=2.6in,angle=-90.]{Figure2A.eps}}\quad
\subfigure{\includegraphics[width=2.6in,angle=-90.]{Figure2B.eps} }}
\caption{The parameter $\eta$ in the fits, shown here in terms of the value of
${\cal R}$ where the first-order diffusion coefficient in ${\cal R}$ is zero (left),
and the value of $\ell$ where the first-order diffusion coefficient in $\ell$ is zero
(right).
Dashed horizontal lines are unweighted means of the plotted points.
\label{Figure:fits12}}
\end{figure}

Figure \ref{Figure:fits12} shows ${\cal R}_0$ as a function of $a$.
There is much scatter, but the mean value is close to ${\cal R}_0=0.5$
($\eta=2$) and there is no obvious trend with $a$.

Based on these results, the following functional forms were adopted:
\begin{subequations}\label{Equation:AdoptedDRDRR}
\begin{eqnarray}
\langle \Delta{\cal R}\rangle_\mathrm{RR} &=& 2 A(E) \left(1 - 2{\cal R}\right), \\
\langle\left(\Delta {\cal R}\right)^2\rangle_\mathrm{RR} &=& 4 A(E) {\cal R}\left(1-{\cal R}\right)  .
\end{eqnarray}
\end{subequations}
Note that the ${\cal R}^2$ term in $\langle\Delta{\cal R}\rangle$ vanishes for
the adopted value of $\eta=2$.
The corresponding diffusion coefficients in $\ell$ are
\begin{subequations}\label{Equation:AdoptedDlDll}
\begin{eqnarray}
\langle\Delta\ell\rangle_\mathrm{RR} &=& A(E) \frac{\left(1-3\ell^2\right)}{2\ell}, \\
\langle\left(\Delta\ell\right)^2\rangle_\mathrm{RR} &=& A(E)\left(1-\ell^2\right)
\end{eqnarray}
\end{subequations}
implying $\langle\Delta\ell\rangle=0$ at $\ell=1/\sqrt{3}\approx 0.58$,
consistent with Figure \ref{Figure:fits12}b.

In what follows, equations (\ref{Equation:AdoptedDRDRR}) are 
assumed to define the RR diffusion coefficients in equations~(\ref{Equation:CombinedDiffCoefs}).

Given these choices, and setting changes in $E$ to zero,
the flux coefficients have the simple forms
\begin{equation}
D_{\cal R}({\cal R}) = 0,\ \ \ \ 
D_{\cal RR}({\cal R})  = 2A(E){\cal R}\left(1-{\cal R}\right)
\end{equation}
and the ${\cal R}-$ directed flux is
\begin{equation}
F_{\cal R} = -2A(E){\cal R}\left(1-{\cal R}\right)\frac{\partial f}{\partial {\cal R}} 
\end{equation}
which has the correct behavior at ${\cal R} = \{0,1\}$.
The zero-flux solution
is therefore $f({\cal R}) = \mathrm{const}$, as in the NR case.
The constant-flux solution, with boundary condition $f=0$ at ${\cal R}={\cal R}_\mathrm{lc}$, is
\begin{eqnarray}
f({\cal R}) = f_0 \log\left[\frac{1-{\cal R}_\mathrm{lc}}{1-{\cal R}}
\frac{\cal R}{{\cal R}_\mathrm{lc}}\right]
\end{eqnarray}
and the flux is $-2A(E) f_0$.
Since
\beq
f(E) \approx \overline{f} \equiv \int f({\cal R}) d{\cal R} = -f_0 \log {\cal R}_\mathrm{lc} ,
\eeq
we can write
\beq
f_0 \approx \frac{f(E)}{\log(1/{\cal R}_\mathrm{lc})}
\eeq
yielding for the flux
\beq
\phi_{\cal R} \approx -\frac{2 A(E) f(E)}{\log(1/{\cal R}_\mathrm{lc})}.
\eeq
We expect these expressions to be only approximate since in reality,
the steady-state solutions will have nonzero fluxes in the $E-$ direction.

We also give here the expression for ${\cal D}$, equation (\ref{Equation:calDofElc}), in the RR regime:
\begin{eqnarray}\label{Equation:DefDRR}
{\cal D} = \left[\frac{\langle\left(\Delta{\cal R}\right)^2\rangle_t}{2{\cal R}}\right]_{{\cal R}_\mathrm{lc}} 
= 2\alpha_s^2\left(\frac{M_\star}{\mh}\right)^2 
\frac{1}{N} \frac{t_\mathrm{coh}}{P^2} \;.
\end{eqnarray}
Using the approximate expression for ${\cal R}_\mathrm{lc}$ in equation (\ref{Equation:CaptureCondApprox}) (corresponding to the capture radius for a compact object),
the quantity $q_\mathrm{lc}$ defined in equation (\ref{Equation:Defineqlc}) becomes
\beq
q_\mathrm{lc} (a) \approx \frac{\alpha_s^2}{8} \left(\frac{m_\star}{\mh}\right)^2 N
\frac{t_\mathrm{coh}}{P} \frac{a}{r_g}.
\eeq
Assuming respectively that $t_\mathrm{coh} = \{t_\mathrm{coh,M}, t_\mathrm{coh,S}\}$
yields
\begin{equation}\label{Equation:qRRMS}
q_\mathrm{lc,M} = \frac{\alpha_s^2}{8} \frac{m_\star}{\mh} \frac{a}{r_g} ,
\ \ \ \ 
q_\mathrm{lc,S} = \frac{\alpha_s^2}{96} \left(\frac{m_\star}{\mh}\right)^2 N(a)
\left(\frac{a}{r_g}\right)^2 .
\end{equation}
Note the interesting result that $q_\mathrm{lc,M}$ does not depend on the mass
distribution, while $q_\mathrm{lc,S}$ does.
The former is
\beq
q_\mathrm{lc,M} \approx 0.07 \left(\frac{\mh/m_\star}{10^5}\right)^{-1} 
\left(\frac{\mh}{10^6\msun}\right)^{-1} \frac{a}{\mathrm{mpc}} .
\eeq

\subsection{Anomalous relaxation}
\label{Section:AR}
The diffusion coefficients defined in the previous section are affected by general
relativity (GR) to the extent that GR determines the coherence time; the latter defined as 
a {\it typical} precession time of all stars at a given radius.
Another GR-related phenomenon 
is the ``Schwarzschild barrier'' (SB), the tendency of {\it single}, very eccentric orbits
not to diffuse in $L$ below some definite value at each $a$.
The SB was first observed in $N$-body simulations \citep{MAMW2011}, 
as a locus in the $\log a$ {\it vs.} $\log(1-e)$ plane where the orbital trajectories ``bounced''
in the course of their RR-driven random walks in $L$.
The same study revealed that orbits experiencing the ``bounce'' were 
of such high eccentricity that their GR precession times were short compared with those
of typical (i.e., higher-$L$) stars at the same $a$.
\citet{Hamers2014} coined the term ``anomalous relaxation'' to describe the
behavior of orbits in this high-eccentricity regime.

Two analytic expressions have been proposed for the location of the SB.
The first is based on a comparison of the GR precession time with the time for the
$\sqrt{N}$ torques to change $L$ \citep{MAMW2011}:
\beq\label{Equation:SB}
{\cal R}_\mathrm{SB}^{(i)} (a) \approx \left(\frac{r_g}{a}\right)^2
\left[\frac{\mh}{M_\star(a)}\right]^2 N(a) \;.
\eeq
The second \citep{Hamers2014} compares the GR precession time with the coherence time:
\beq\label{Equation:SB2}
{\cal R}_\mathrm{SB}^{(ii)} (a) \approx 4 \frac{r_g}{a} \frac{t_\mathrm{coh}(a)}{P(a)}\;.
\eeq
In spite of their disparate functional forms, the two expressions yield  numerically
similar relations for ${\cal R}_\mathrm{SB}(a)$ in a wide range of nuclear models.
However the former relation appears to more
accurately reproduce the barrier location in numerical studies to date \citep{MAMW2011,Hamers2014}.

Appendix C derives expressions for the
diffusion coefficients in this regime, based on a simple Hamiltonian model 
\citep{MAMW2011}:
\begin{equation}\label{Equation:DiffARapprox}
\langle\Delta\ell\rangle \approx \frac{2\ell^3}{\tau}  , \ \ \ \ 
\langle\left(\Delta\ell\right)^2\rangle \approx \frac{\ell^4}{\tau}  
\end{equation}
where $\tau(a) \equiv t_\mathrm{coh}(a)/(A_\mathrm{\sqrt{N}})^2$
and
\beq\label{Equation:DefineAD}
A_{\sqrt{N}} \equiv \frac{1}{2\sqrt{N(a)}}
\frac{M_\star(a)}{\mh} \frac{a}{r_g} .
\eeq
\citet{Hamers2014} verified the $\ell$-dependence predicted by equation (\ref{Equation:DiffARapprox})
via numerical experiments.
Expressed in terms of ${\cal R}$ via equations (\ref{Equation:elltoR}), 
the diffusion coefficients become
\begin{eqnarray}
\langle\Delta{\cal R}\rangle_\mathrm{AR} =
\frac{5}{\tau}  {\cal R}^2, \ \ \ \ 
\langle(\Delta{\cal R})^2\rangle_\mathrm{AR} 
= \frac{4}{\tau} {\cal R}^3 .
\label{Equation:DiffAR12}
\end{eqnarray}
Equations (\ref{Equation:DiffAR12}) replace equations (\ref{Equation:AdoptedDRDRR}) when
${\cal R} \le {\cal R}_\mathrm{SB}$.
We note that the values of the numerical coefficients in these expressions have not been well determined
by the numerical experiments to date,
and the simple model in Appendix C is not likely to be valid when ${\cal R}$ is so
small that the GR precession time is much shorter than the coherence time.
Additional $N$-body experiments are needed to elucidate the functional forms of the anomalous
diffusion coefficients.
Solutions to the Fokker-Planck equation containing these diffusion coefficients will be postponed
to Paper III in this series \citep{PaperIII}.

\bigskip
\subsection{Gravitational radiation}
First-order diffusion coefficients in ${\cal E}$ and ${\cal R}$ also exist that describe
changes due to emission of gravitational waves.
At the lowest (2.5 PN) order, the orbit-averaged rates of change of $a$ and $e$ are:
\begin{subequations}
\begin{eqnarray}
\langle\Delta a\rangle &=& -\frac{64}{5} \frac{G^3\mh^2 m_\star}{c^5 a^3\left(1-e^2\right)^{7/2}}\left(1+\frac{73}{24}e^2 + \frac{37}{96} e^4\right) , \\
\langle\Delta e\rangle &=& -\frac{304}{15} \frac{G^3\mh^2 m_\star e}{c^5 a^4\left(1-e^2\right)^{5/2}}\left(1+\frac{121}{304}e^2\right)
\end{eqnarray}
\end{subequations}
(equations 4.234 from \citet{DEGN} after replacing $m_1m$ in those expressions by $\mh^2$).
Expressed in terms of ${\cal E}=G\mh/(2a)$ and ${\cal R}=1-e^2$,
\begin{subequations}
\begin{eqnarray}
\langle \Delta {\cal E}\rangle &=& -\frac{2{\cal E}^2}{G\mh}\langle\Delta a\rangle 
= \frac{2720}{3} \frac{m_\star}{\mh} \frac{1}{G\mh c^5}
\frac{{\cal E}^5}{{\cal R}^{7/2}} \left(1-\frac{366}{425}{\cal R} + \frac{37}{425}{\cal R}^2\right)
, \\
\langle \Delta {\cal R}\rangle &=& -2\sqrt{1-{\cal R}}\langle\Delta e\rangle =
\frac{2720}{3} \frac{m_\star}{\mh} \frac{1}{G\mh c^5}
\frac{{\cal E}^4}{{\cal R}^{5/2}} \left(1-\frac{546}{425}{\cal R} + \frac{121}{425}{\cal R}^2\right).
\end{eqnarray}
\end{subequations}
\bigskip
\section{Numerical algorithm}
\subsection{Choice of grid and units}
\begin{figure}[h]
\begin{center}
  \includegraphics[angle=-90.,width=7.0in]{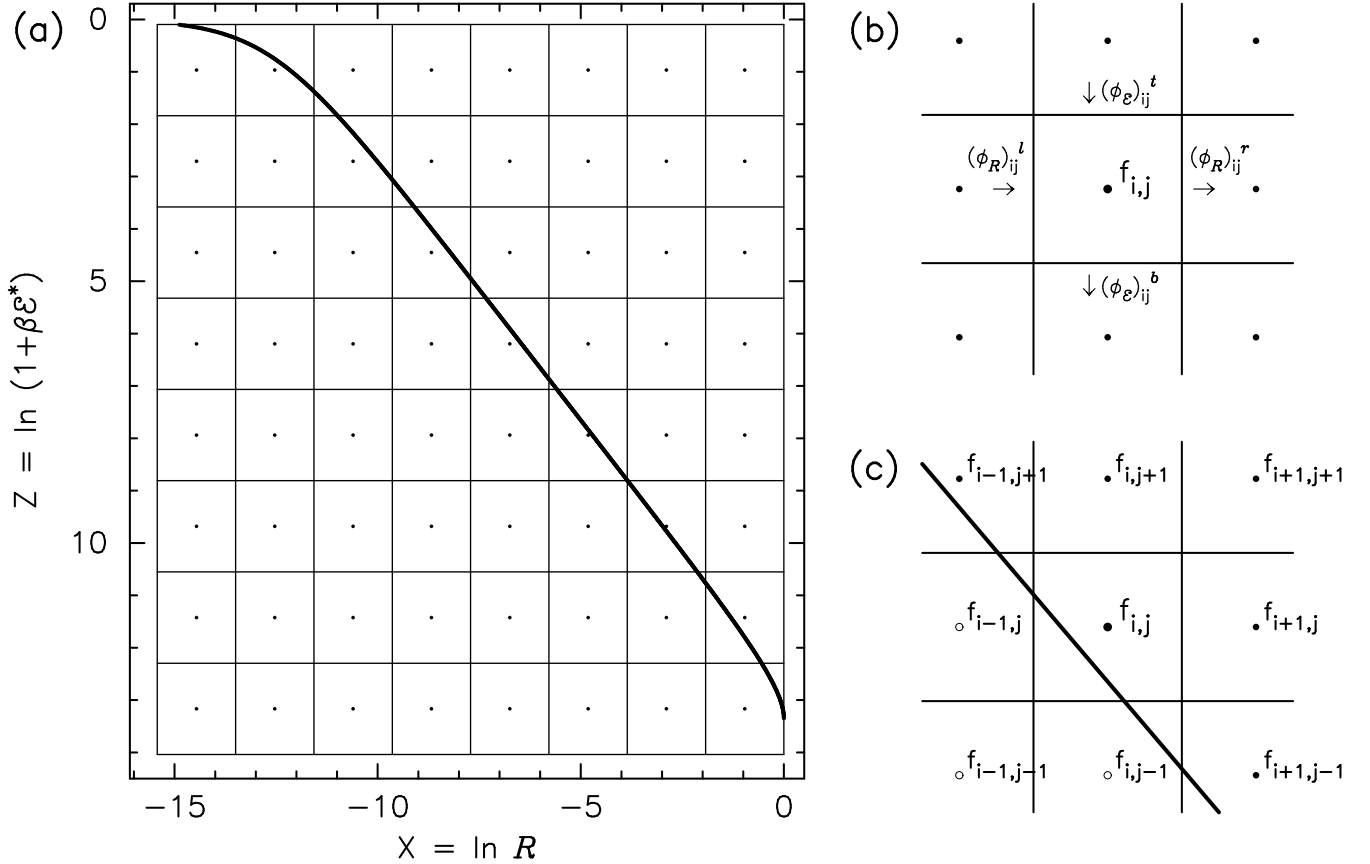}
  \caption{(a) Solution grid; dots denote cell centers where the $f_{ij}$ values are defined.
  Fluxes are defined on the cell boundaries. Thick solid curve is the boundary of the
  loss cone, ${\cal R} = {\cal R}_\mathrm{lc}({\cal E})$. 
  (b) Schematic diagram showing the fluxes associated with grid cell ($i,j$).
  (c) Finite-differencing for grid cells adjacent to (above) the loss cone boundary
  is affected by the lack of knowledge of $f$ inside the loss cone. In the case of 
  ``empty loss cone'' boundary conditions, shown here, $f$ is assumed to be zero
   inside the loss cone (open circles). Cohn-Kulsrud boundary conditions are expressed 
   purely in terms of the values of $f$ near the loss-cone boundary; no derivatives
   are evaluated there and the values of $f$ inside the loss cone are not needed.}
\label{Figure:grids}
\end{center}
\end{figure}

Solutions are obtained on a ($N_x \times N_z$) grid in $(X,Z)$, where
\begin{subequations}
\begin{eqnarray}
X &\equiv & \ln R = \ln \left[\frac{L}{L_c({\cal E})}\right]^2  = \ln \left(1-e^2\right), \\
Z &\equiv & \ln\left(1 + \beta {\cal E}^*\right) = \ln \left(1+\beta {\cal E}/c^2\right)
\end{eqnarray}
\end{subequations}
(Figure~\ref{Figure:grids}).
Typically $N_x=N_z=64$.
Here and below, starred variables are dimensionless, based on code units which are
\beq
G = \mh = c = 1 .
\eeq
Thus ${\cal E}=c^2 {\cal E}^\star$ and $L=(G\mh/c)L^\star$.
The angular momentum of a circular orbit,
$L_c = \sqrt{G\mh/(2{\cal E})}$, becomes
in dimensionless units $L_c^* =1/(2{\cal E}^*)$.
The squared angular momentum of an orbit with periapsis at $r=8r_g$ is
\beq
R_\mathrm{lc}({\cal E}^*) = 32 {\cal E}^* \left(1-8{\cal E}^*\right), \ \ {\cal E}^*\le \frac{1}{16}\; .
\eeq

In the code, all diffusion coefficients are divided by the
factor $\Gamma\equiv 4\pi G^2m_\star^2\ln\Lambda$.
The dimensional factor $f_0$ multiplying $f^*$
is taken to be
\beq
f_0 = \left(\frac{r_m}{r_g}\right)^\gamma \left(\frac{\mh}{m_\star}\right) \frac{1}{c^3r_m^3} .
\eeq
These two choices determine the unit of time as $(\Gamma f_0)^{-1}$ or
\begin{eqnarray}\label{Equation:Definet0}
[t] \equiv t_0 = \frac{1}{4\pi\ln\Lambda} \left(\frac{\mh}{m_\star}\right) 
\left(\frac{r_m}{r_g}\right)^{2-\gamma} \left(\frac{r_m}{c}\right) 
= 1.7\times 10^{-4} \mathrm{yr} \left(\frac{\ln\Lambda}{15}\right)^{-1}
\left(\frac{\mh/m_\star}{10^5}\right) \left(\frac{r_m}{r_g}\right)^{3-\gamma}\; .
\end{eqnarray}
In these expressions,  $r_m$ and $\gamma$ are free parameters, but they
have a simple physical interpretation in the case of a cluster with a power-law
density:
\begin{eqnarray}\label{Equation:nofrpowerlaw}
n(r) = n_0\left(\frac{r}{r_m}\right)^{-\gamma},\ \ \ \ 
n_0 = \frac{3-\gamma}{2\pi} \frac{\mh}{m_\star} \frac{1}{r_m^3} 
\end{eqnarray}
for which  $r_m$ is the radius containing a mass in stars equal to
$2\mh$.
The isotropic $f$ corresponding to the density (\ref{Equation:nofrpowerlaw}) is
\begin{eqnarray}\label{Equation:fofE}
f({\cal E}) = f_0 f^*({\cal E}^*) = 
f_0C(\gamma) {{\cal E}^\star}^{\gamma-3/2}, \ \ \ \ 
C(\gamma) = \frac{3-\gamma}{8}\sqrt{\frac{2}{\pi^5}} \frac{\Gamma(\gamma+1)}{\Gamma(\gamma-1/2)}\; .
\end{eqnarray}
Initial conditions were typically chosen to be equations (\ref{Equation:nofrpowerlaw}) --
(\ref{Equation:fofE}) or some modification, e.g., $f$ might be set initially to zero inside the
loss cone, as described in more detail below.
When the initial $f$ was constructed in this way, 
$r_m$ and $\gamma$ retain their meaning as the parameters describing the unmodified
power-law model.

The initial density normalization (stars per unit volume) is fixed by the choice of $r_m^* = r_m/r_g$.
At any later time, the dimensionless number density:
\beq\label{Equation:Definenstarofr}
n^*(r^*) = \int f^* d^3v^*
\eeq
is related to the true number density by
\begin{eqnarray}
n(r) = \int f d^3 v = c^3 f_0 \int f^* d^3 v^* 
= \frac{1}{r_g^3} \left(\frac{r_m}{r_g}\right)^{\gamma-3} \frac{\mh}{m_\star} \; n^*(r^*).
\end{eqnarray}
Similarly the number of stars within $r$ is
\beq\label{Equation:NofrNstar}
N(<r) = \left(\frac{r_m}{r_g}\right)^{\gamma-3} \left(\frac{\mh}{m_\star}\right) N^*(<r^*)
\eeq
where $N^*(r^*) = 4\pi \int_0^{r^*} n^*(r^*) {r^*}^2 dr^*.$ 
These expressions are used when evaluating the diffusion coefficients in the RR and AR regimes.
Combining equations (\ref{Equation:Definet0}) and (\ref{Equation:NofrNstar}),
the rate of loss of stars from within a sphere of radius $r$  is given in terms of dimensionless 
quantities as 
\begin{eqnarray}\label{Equation:DefineNdot}
\frac{dN(<r)}{dt} &=& 4\pi\ln\Lambda \frac{c}{r_g} \left(\frac{r_m}{r_g}\right)^{2(\gamma-3)} 
\frac{dN^*(<r^*)}{dt^*} .
\end{eqnarray}

The dimensionless function $q_\mathrm{lc}({\cal E})$, equation (\ref{Equation:Defineqlc}),
that appears in the Cohn-Kulsrud boundary layer treatment also depends on $m_\star/\mh$.
Since
\beq
{\cal D}({\cal E}) = \Gamma f_0 {\cal D}^*({\cal E}^*),
\eeq
we can write
\beq\label{Equation:qlcofEstar}
q_\mathrm{lc}({\cal E}^*) = 2\sqrt{2}\pi^2\ln\Lambda \frac{m_\star}{\mh} \left(\frac{r_m}{r_g}\right)^{\gamma-3} \frac{{\cal D}^*({\cal E}^*)}{{{\cal E}^*}^{3/2}} \frac{1}{{\cal R}_\mathrm{lc}}
\eeq
where
\beq
{\cal D}^*({\cal E}^*) = \frac{D_{\cal RR}^*\left({\cal E}^*, {\cal R}_\mathrm{lc}\right)}{{\cal R}_\mathrm{lc}}
\eeq
is the dimensionless form of the function defined in equation (\ref{Equation:calDofElc}).
For instance, in the case that diffusion in $L$ is due only to non-resonant relaxation,
equations (\ref{Equation:qlcofEstar}) and (\ref{Equation:qlcNR}) imply
\beq\label{Equation:qlcofFstar}
q_\mathrm{lc}({\cal E}^*) = \frac{2\sqrt{2}\pi^2}{3}\frac{m^\star}{\mh} \ln\Lambda
\left(\frac{r_m}{r_g}\right)^{\gamma-3} 
\frac{5F_0^* + 12F_3^* - 4F_7^*} 
{{{\cal E}^*}^{5/2} {\cal R}_\mathrm{lc}}.
\eeq 
\subsection{Basic numerical approach}
As in \citet{CohnKulsrud1978}, the distribution function $f$ is evaluated at the cell centers
while the flux coefficients and fluxes are evaluated at the cell boundaries (Figure \ref{Figure:grids}a).
The rate of change of ${\cal J}f$ in a cell is then equated to the net flux through the cell boundaries
(Figure \ref{Figure:grids}b).
The dimensionless evolution equation:
\begin{equation}
\frac{\partial f^\star}{\partial t^\star} = -{{\cal J}^\star}^{-1} \beta e^{-Z} \frac{\partial}{\partial Z} \left({\cal J}^\star\phi_{\cal E}^\star\right) - e^{-X} \frac{\partial}{\partial X} \phi_{\cal R}^\star
\end{equation}
was  written in implicit discrete form, with the
time derivative represented as a backwards difference and the flux coefficients evaluated
at the previous time step.
In discrete form, the $N_X\times N_Z$ difference equations become
\begin{eqnarray}\label{Equation:Matrix}
\frac{f_{ij}^{(n+1)} - f_{ij}^{(n)}}{\Delta t} &=& \sum_{k=i-1}^{i+1} \sum_{l=j-1}^{j+1} A_{ijkl} f_{kl}^{n+1}
\end{eqnarray}
where $f_{ij}^{(n)} = f(Z_i,X_j)$ evaluated at the $n$th time step.
Solution of the matrix equation (\ref{Equation:Matrix}) is carried out using the fortran program {\tt f07acf} from {\it Numerical Algorithms Group}.
This routine uses iterative refinement to produce a solution with double-precision accuracy.
Iteration was found to be unstable when the time step, $\Delta t$, was taken to be too large; the maximum allowable $\Delta t$ was found to decrease with increasing number of grid cells.
Less accurate matrix inversion routines were found to be faster but also more susceptible to instability.

\cite{CohnKulsrud1978} employed a linear grid in ${\cal R}$ extending from ${\cal R}=0$
to ${\cal R}=1$.
As a result, the dependence of $f$ on ${\cal R}$ near the loss cone was not well
resolved.
The use here of a logarithmic grid in ${\cal R}$ yields many more grid points at small ${\cal R}$,
permitting a more careful treatment of the solution near the loss cone, as discussed below.
\subsection{Boundary conditions}
Two choices for the the boundary condition at small ${\cal R}$ are implemented.
The simplest is an ``empty loss cone'' (ELC) condition: $f({\cal E}, {\cal R})=0$
for ${\cal R}\le {\cal R}_\mathrm{lc}({\cal E})$.
To implement this condition, cells that contain the curve ${\cal R}= {\cal R}_\mathrm{lc}({\cal E})$
and for which the cell center is outside the loss cone (``loss-cone cells'') are identified, and for these cells, a record is made of the adjacent cells for which the cell center
is inside the loss cone, i.e. for which ${\cal R}_{ij}<{\cal R}({\cal E}_{ij})$.
There are nine possible combinations of adjacent cells for which this condition can be 
satisfied; Figure \ref{Figure:grids}c illustrates the most common, consisting of three adjacent
cells.
 The finite-difference expressions for the flux derivatives in the loss-cone cells
 are written with the appropriate $f_{ij}$-values set to zero.

The second choice of small-${\cal R}$ boundary condition was based on the Cohn-Kulsrud
(1978)  boundary layer solution.
In the spirit of that derivation, the evolution equations near the loss cone boundary are first simplified
by ignoring the contribution of gradients in ${\cal E}$ to the flux, which allows equations (\ref{Equation:FPFluxConserve}) to be written as
\begin{eqnarray}\label{Equation:FluxLC}
\phi_{\cal E} = -D_{\cal E\cal R} \frac{\partial f}{\partial {\cal R}} -
D_{\cal E} f,  \ \ \ \ 
\phi_{\cal R} = -D_{\cal R\cal R} \frac{\partial f}{\partial {\cal R}} - D_{\cal R} f .
\end{eqnarray}
In the Cohn-Kulsrud solution, the ${\cal R}$-directed
flux per unit of ${\cal E}$ across the loss cone boundary:
\begin{eqnarray}
F({\cal E}) = -\int_{{\cal R}_\mathrm{lc}}^1 \left(-{\cal J}\frac{\partial\phi_{\cal R}}{\partial {\cal R}}\right)d{\cal R} 
= {\cal J} ({\cal E}) \left[\phi_{\cal R}({\cal R}=1) - \phi_{\cal R}({\cal R} = {\cal R}_\mathrm{lc})\right]
= -{\cal J}({\cal E}) \phi_{\cal R}({\cal R}_\mathrm{lc}) 
\end{eqnarray}
is given by
\begin{eqnarray}\label{Equation:InternalFlux}
F({\cal E}) = 4\pi^2L_c^2({\cal E})\; {\cal R}_\mathrm{lc}({\cal E}) \; f({\cal E},{\cal R}_\mathrm{lc})\; \xi(q_\mathrm{lc})
\end{eqnarray}
where
\beq\label{Equation:Definexiofq}
\xi(x) \equiv 1 - 4\sum_{m=1}^\infty\frac{e^{-\alpha_m^2x/4}}{\alpha_m^2}
\eeq
and the $\alpha_m$ are the consecutive zeros of the Bessel function $J_0(\alpha)$
\citep[][equations 6.58-6.62]{DEGN}.
Thus
\begin{eqnarray}\label{Equation:FluxLCR}
\phi_{\cal R}({\cal E}, {\cal R}_\mathrm{lc}) = -\frac{\sqrt{2}}{\pi} \frac{{\cal E}^{3/2}}{G\mh} {\cal R}_\mathrm{lc}({\cal E}) \; f_\mathrm{lc}({\cal E}) \; \xi(q_\mathrm{lc})
\end{eqnarray}
where $f_\mathrm{lc}({\cal E}) \equiv f({\cal E}, {\cal R}_\mathrm{lc})$.
The gradient in $f$ at ${\cal R}={\cal R}_\mathrm{lc}$ in the Cohn-Kulsrud solution is
\beq\label{Equation:dfdRlc}
\left(\frac{\partial f}{\partial {\cal R}}\right)_\mathrm{lc} =
\frac{\xi}{q_\mathrm{lc}} \frac{f_\mathrm{lc}}{{\cal R}_\mathrm{lc}}
\eeq
which is inserted into the first of equations (\ref{Equation:FluxLC}) to give the flux in ${\cal E}$:
\begin{eqnarray}\label{Equation:FluxLCE}
\phi_{\cal E}({\cal E}, {\cal R}_\mathrm{lc}) =
-D_{\cal E\cal R}({\cal E}, {\cal R}_\mathrm{lc}) \frac{\xi}{q_\mathrm{lc}} 
\frac{f_\mathrm{lc}({\cal E})}{{\cal R}_\mathrm{lc}({\cal E})}  
- D_{\cal E}({\cal E}, {\cal R}_\mathrm{lc}) f_\mathrm{lc}({\cal E}) .
\end{eqnarray}
Equations (\ref{Equation:FluxLCR}) and (\ref{Equation:FluxLCE}) are the adopted
boundary conditions.
These expressions are applied at the inner edges of any loss-cone cell, while
the general expressions (\ref{Equation:FPFluxConserve}) for the flux 
are applied on the outer edges of those cells.

\begin{figure}[h]
\begin{center}
  \includegraphics[width=3.5in]{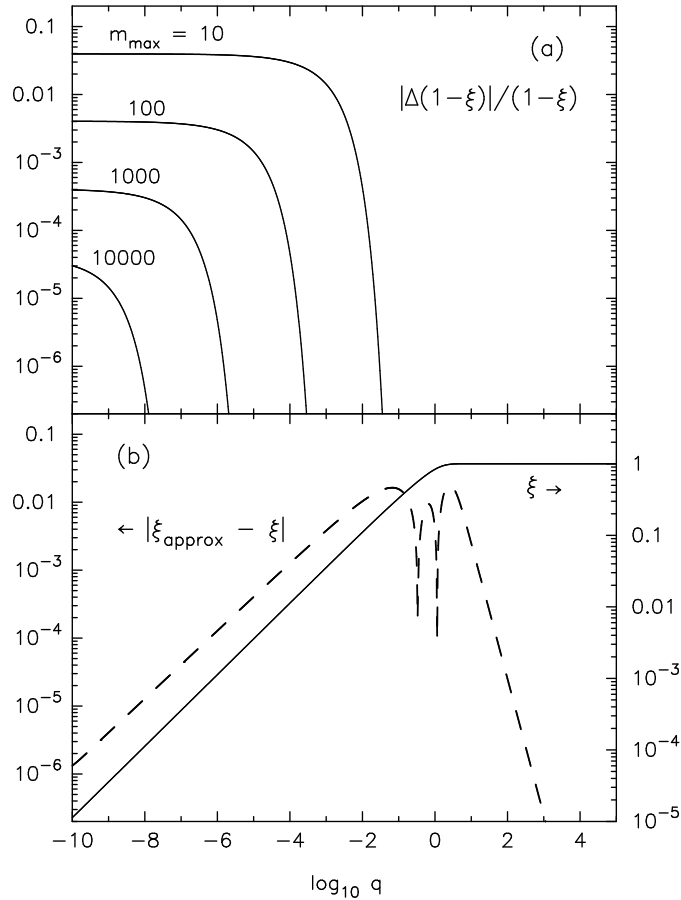}
  \caption{(a)  Convergence of the series in equation (\ref{Equation:Definexiofq}).
  Each curve is labelled by $m_\mathrm{max}$, the number of terms included in the
  summation; the ordinate is the relative error.  (b) Solid curve shows $\xi(q)$ computed
  from equation (\ref{Equation:Definexiofq}) using $10^5$ terms. Dashed curve is
  the absolute error in the approximate expression (\ref{Equation:xiapprox}).}
\label{Figure:xi}
\end{center}
\end{figure}

The summation in equation (\ref{Equation:Definexiofq}) is slowly converging when
$q$ is small (Figure \ref{Figure:xi}a).
An approximate expression is
\beq\label{Equation:xiapprox}
\xi_\mathrm{approx}(x) = \frac{x}{\left(x^2 + x^4\right)^{1/4}}
\eeq
for which the relative error is $\sim 1\%$ at all $q\lesssim 1$ and 
tends to zero for large $q_\mathrm{lc}$ (Figure \ref{Figure:xi}b).
Equation (\ref{Equation:xiapprox}) was adopted for the code.

Note that the values $f_{ij}$ corresponding to grid cells inside the loss cone are not used and
therefore are not advanced in time.
Indeed in this region, $f$ is not expected to satisfy Jeans's theorem and a proper 
treatment would require inclusion of $r-$ as an additional independent variable.

The boundary condition at ${\cal R}=1$ is $\phi_{\cal R} = 0$.

Two possibilities were implemented for the boundary condition at small ${\cal E}$, i.e. far from the
\sbh.
The first consists of fixing $f({\cal E}_\mathrm{min}, {\cal R})$ at its initial value.
\citet{CohnKulsrud1978} implemented a similar boundary condition.
This choice implies a nonzero flux of stars into the region of integration, 
although that flux can be very small if ${\cal E}_\mathrm{min}$ is small.
The second choice consists of setting to zero the
${\cal E}-$ directed flux at ${\cal E} = {\cal E}_\mathrm{min}$,
i.e. requiring 
\beq\label{Equation:ZeroBoundary}
D_{\cal EE} \frac{\partial f}{\partial {\cal E}} = -D_{\cal ER} \frac{\partial f}{\partial {\cal R}}
- D_{\cal E} f, \ \ {\cal E} = {\cal E}_\mathrm{min} .
\eeq
This boundary condition is more in keeping with $N$-body simulations that have
a fixed number of stars.
The ``zero-flux'' boundary condition is implemented as follows.
(1) At each time step, inversion of the matrix $A_{ijkl}$ is first carried out
specifying no change at ${\cal E}={\cal E}_\mathrm{min}$, i.e., $Z=Z_{NZ}$.
(2) Equation (\ref{Equation:ZeroBoundary}) is then used to solve for 
$f(Z_{NZ}, X_j), j = 1, \ldots, NX$. In this step, derivatives are expressed
in terms of the unknown values of $f$ at $Z_{NZ}$ and the just-computed values at
$Z_{NZ-1}$, allowing solution of $f$ by inversion of a tri-diagonal matrix.
\subsection{Initial conditions}
 The initial $f({\cal E},{\cal R})$ was typically based on an isotropic power-law model, 
 equation (\ref{Equation:fofE}), 
 but with a modified ${\cal R}$-dependence to account for the presence of the loss cone.
The simplest choice is to set $f=0$ for ${\cal R}\le{\cal R}_\mathrm{lc}$.
A more natural choice is to set the initial $f$ to
\begin{eqnarray}\label{Equation:finitlog}
f({\cal E}, {\cal R}) &=& 
f({\cal E},1)\frac{\ln({\cal R}/{\cal R}_\mathrm{lc})}{\ln(1/{\cal R}_\mathrm{lc})},\ \ \ \ 
{\cal R} > {\cal R}_\mathrm{lc} \nonumber \\
&=& 0, \ \ \ \ {\cal R} \le {\cal R}_\mathrm{lc}
\end{eqnarray}
 which is the approximate (small-${\cal R}$) steady-state solution for an empty loss cone.
 Other choices for the initial conditions are described below.

\section{Examples}
Subsequent papers in this series will describe the evolution of $f(E,L)$ due
to the various physical mechanisms represented above.
Here we use the algorithm to explore some simple dependencies of the classical
Bahcall-Wolf solution on $m_\star/\mh$, and on time.

\subsection{Models with classical diffusion coefficients; empty loss cone}
The number of parameters that define an integration can be minimized by assuming 
empty-loss-cone (ELC) boundary conditions, and adopting the classical (CK) diffusion coefficients.
The stellar mass, $m_\star$, then appears only in expressions like (\ref{Equation:Definet0}) 
that set the scaling of the time or density, and any initial $f({\cal E}, {\cal R})$ 
should evolve to the same equilibrium $f$ modulo a known scaling in amplitude.
Furthermore this steady state should be similar to the Bahcall-Wolf ``zero-flux'' model,
$n\sim r^{-7/4}$, $f\sim {\cal E}^{1/4}$; the differences being due to the fact that 
those authors assumed $f=f({\cal E}, t)$ and allowed loss of stars to the \sbh\ 
only via diffusion in energy.

The code was tested under these assumptions, starting from power-law models like 
those of equation~(\ref{Equation:fofE}) and various values of $\gamma$.
The isotropic models were modified initially as in equation (\ref{Equation:finitlog}) 
so that $f$ fell to zero at ${\cal R}={\cal R}_\mathrm{lc}$; 
the ELC boundary conditions then guaranteed that $f({\cal R}_\mathrm{lc})$ 
remained zero at all ${\cal R}\le{\cal R}_\mathrm{lc}$.
$f({\cal E}_\mathrm{min}, {\cal R})$ was fixed to its initial form.

It is permissible and convenient to scale these models by assuming that the final
mass density has a specified value at a specified radius.
Let that radius be $\alpha r_g$ and let the density at that radius be $\rho(\alpha r_g)$.
It may be shown, using equation (\ref{Equation:Definet0}), 
that the elapsed time is then given in terms of the dimensionless time $t^*$ by
\begin{eqnarray}
t &=& \frac{1}{4\pi\ln\Lambda} \left(\frac{\mh}{m_\star}\right) \left(\frac{r_g}{c}\right)
\left[\frac{\rho(\alpha r_g) r_g^3}{\mh}\right]^{-1} n^*(\alpha)\; t^* \nonumber \\
&\approx& 7.4\times 10^{17} \left(\frac{\ln\Lambda}{15}\right)^{-1} \left(\frac{\mh/m_\star}{10^5}\right)
\left(\frac{\mh}{10^6\msun}\right)^{-1} \left[\frac{\rho(\alpha r_g)}{10^6 \msun \mathrm{pc}^{-3}}\right]^{-1} n^* (\alpha) \; t^* \
\mathrm{yr}
\label{Equation:tscaleELC}
\end{eqnarray}
where $n^*(\alpha)$ is the final, dimensionless model density (cf. equation \ref{Equation:Definenstarofr}) at $r=\alpha r_g$,
i.e. at $r^*=\alpha$.
The loss rate, equation (\ref{Equation:DefineNdot}), can similarly be related to the dimensionless
loss rate as
\begin{eqnarray}
\frac{dN}{dt} &=& 4\pi\ln\Lambda \left(\frac{c}{r_g}\right) \left[\frac{\rho(\alpha r_g) r_g^3}{\mh^2}\right]^2 \frac{1}{n^*(\alpha)^2} \frac{dN^*}{dt^*} \nonumber \\
&\approx& 1.49\times 10^{-35} \left(\frac{\ln\Lambda}{15}\right)
\left(\frac{\mh}{10^6\msun}\right)^3 \left[\frac{\rho(\alpha r_g)}{10^6\msun \mathrm{pc}^{-3}}\right]^2
\frac{1}{n^*(\alpha)^2} \frac{dN^*}{dt^*} \; \mathrm{yr}^{-1} .
\label{Equation:NdotscaleELC}
\end{eqnarray}

\begin{figure}[h!]
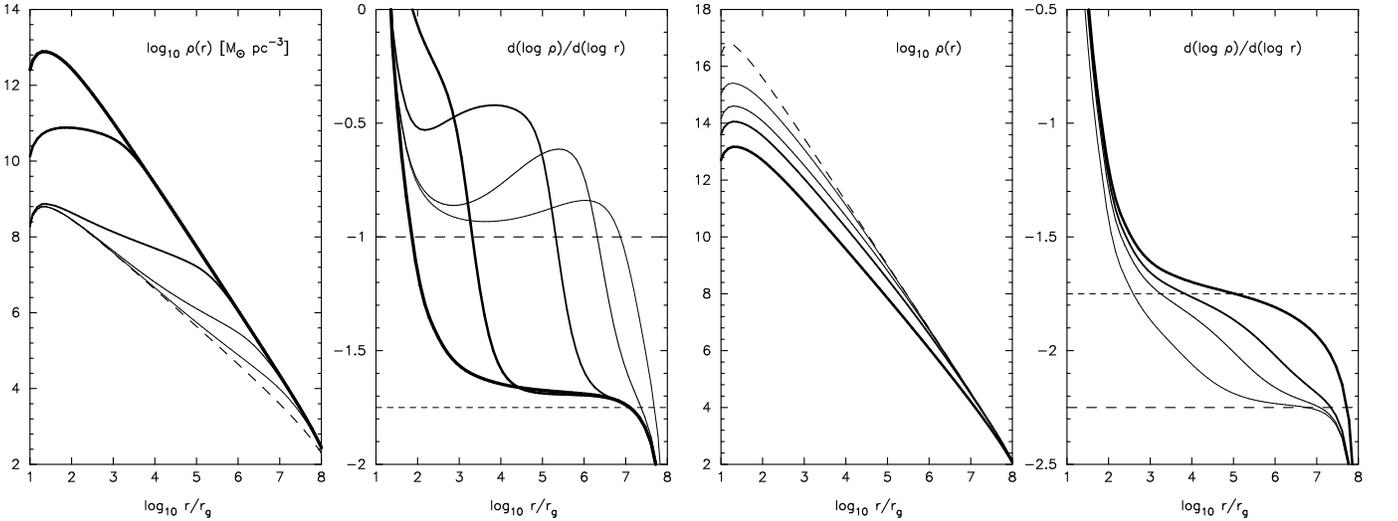

\centering
\mbox{\subfigure{\includegraphics[angle=-90.,width=3.5in]{Figure5A.eps}}\quad
\subfigure{\includegraphics[angle=-90.,width=3.5in]{Figure5B.eps} }}
\caption{Evolution due to classical relaxation of two models assuming empty-loss-cone (ELC) boundary
  conditions. The initial model had $n\sim r^{-1}$ (left) and $n\sim r^{-9/4}$ (right),
  shown as the dashed lines.
  Other curves are at times $(1,2,3,4,5)\times 10^9$ yr (left)
  and $(0.01,0.1,0.5,7) \times 10^8$ yr (right) based on the scaling described in the text; line thickness
  increases with time.
  Dotted lines show $d\log\rho/d\log r=-7/4$, the  Bahcall-Wolf slope.}
 \label{Figure:denrun11}
\end{figure}

Figure \ref{Figure:denrun11} shows the evolution of $\rho(r)$ for integrations with 
 $\gamma=1$ and $\gamma=9/4$.
The units of time and density were fixed by assuming a final density at $10^6 r_g$ of
$10^6\msun \mathrm{pc}^{-3}$; the other parameters were set to the fiducial values
in equations (\ref{Equation:tscaleELC}) and (\ref{Equation:NdotscaleELC}), 
e.g. $\mh=10^6\msun$, hence $10^6r_g \approx 4.8\times 10^{-2}$ pc; 
the radius containing a mass in stars of $2\mh$ at the final time step, 
the ``gravitational influence radius,'' is $\sim 3\times 10^8 r_g$, roughly equal to the
outer radius of the solution grid. 
In the model with initially shallower slope ($\gamma=1$),  the density evolves 
``from the outside in'' toward the steady state, while in the model with $\gamma=9/4$
the density evolves at all radii at roughly the same rate.

The final density profile is the same in these two models, as expected, and is close to a
power law:
\beq
\rho(r) \sim r^{-\delta}, \ \ \ \ 1.65\lesssim \delta \lesssim 1.70
\eeq
at $10^4 r_g\le r \le 10^7 r_g$.
This is slightly shallower than the canonical $\rho\sim r^{-1.75}$ of the Bahcall-Wolf solution.
The difference can be attributed to the depletion of orbits in the loss cone, the effects of which
are  progressively more severe at energies close to the \sbh.

\begin{figure}[h]
\begin{center}
  \includegraphics[angle=-90.,width=7.0in]{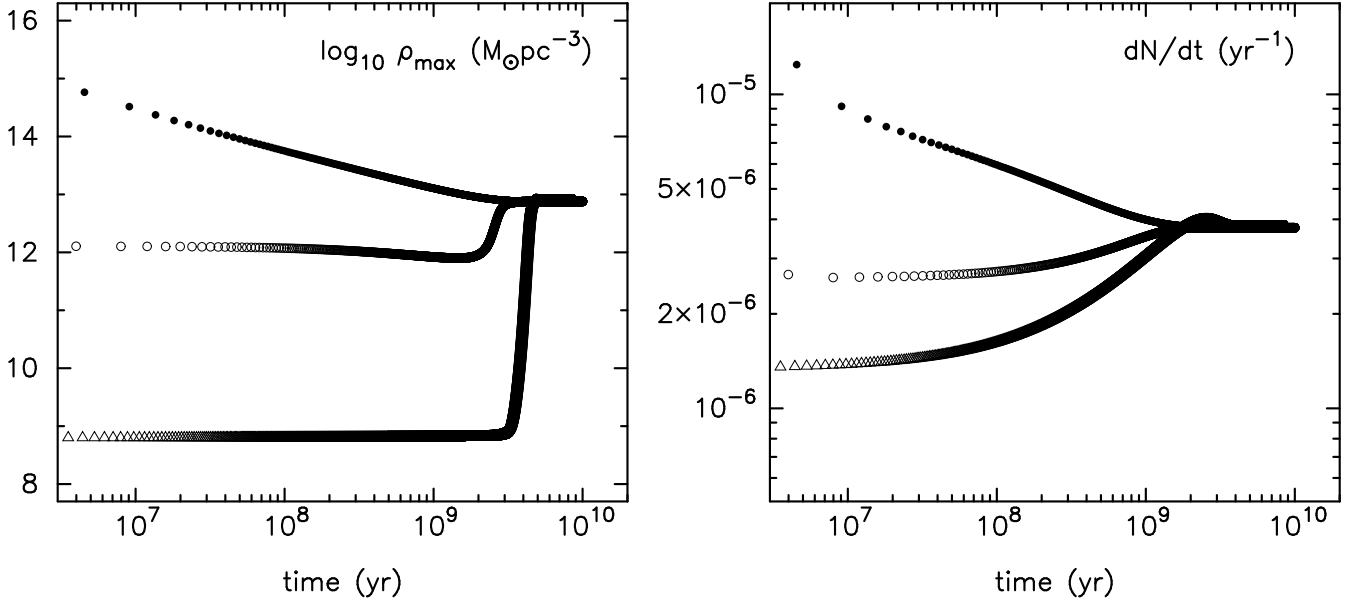}
  \caption{Evolution of the peak density (left) and of the total loss rate (right) for three
  integrations assuming classical diffusion coefficients and ELC boundary conditions.
  The initial density slope was $\gamma=1$ (triangles), $\gamma=1.5$ (open circles),
  and $\gamma=2$ (filled circles).
  Scaling assumes a final density at $10^6r_g$ of $10^6\msun$ pc$^{-3}$.}
\label{Figure:mndot}
\end{center}
\end{figure}

Figure \ref{Figure:mndot} shows the evolution of the peak ($\sim$ central) density, 
as well as the total loss rate $\dot N$, for three integrations starting from
$\gamma=\{1,3/2,2\}$.
In each case a steady state is reached in a time of $\sim 5 \times 10^9$ yr.
This can be compared to the classical relaxation time, a 
standard expression for which is
\begin{eqnarray}\label{Equation:DefineTr}
&&T_r(r) = {0.34\sigma^3\over G^2 m\rho\ln\Lambda} \\
&&\approx
0.95\times 10^{10} \!\left({\sigma\over 200\,\mathrm{km\,s}^{-1}}\right)^{\!3} \!\!\left({\rho\over 10^6\,\msun\,\mathrm{pc}^{-3}}\right)^{\!-1} \!\!\left({m_\star\over \msun}\right)^{\!-1} \!\!\left({\ln\Lambda\over 15}\right)^{\!-1}\!\mathrm{yr} \nonumber
\end{eqnarray}
 \cite[][Eq. 3.1]{DEGN}.
 Evaluating equation (\ref{Equation:DefineTr}) in the {\it final} model, 
 assuming $n(r)\sim r^{-1.7}$, $m_\star=10^{-5}\mh=10\msun$, $\ln\Lambda = 15$,  and
\beq
\sigma(r) \approx \left[\frac{1}{1+\gamma} \frac{G\mh}{r}\right]^{1/2} \approx
183 \left(\frac{r}{10^6 r_g}\right)^{-1/2} \mathrm{km\ s}^{-1}, 
\eeq
yields
\beq
T_r(r) \approx 7\times 10^9 \left(\frac{r}{10^6 r_g}\right)^{0.20} \mathrm{yr},
\eeq
consistent with the observed equilibration time.
\subsection{Models with classical diffusion coefficients; Cohn-Kulsrud loss cone}
The simple rescaling defined in the previous section no longer works when 
adopting the Cohn-Kulsrud (CK) loss-cone boundary conditions,
equations (\ref{Equation:FluxLCR}) -- (\ref{Equation:FluxLCE}),
since the dimensionless function $q_\mathrm{lc}({\cal E},t)$ 
depends separately on the factor
\beq\label{Equation:qterm}
\frac{m_\star}{\mh}\ln\Lambda
\eeq
(equations \ref{Equation:Defineqlc}, 
\ref{Equation:qlcofFstar}).
Two models with different initial $f$'s will not necessarily arrive at steady states
that are rescaled versions of one another
unless the final $q_\mathrm{lc}({\cal E})$ happens also to be the same.

\begin{figure}[h]
\begin{center}
  \includegraphics[angle=90.,width=8.75in]{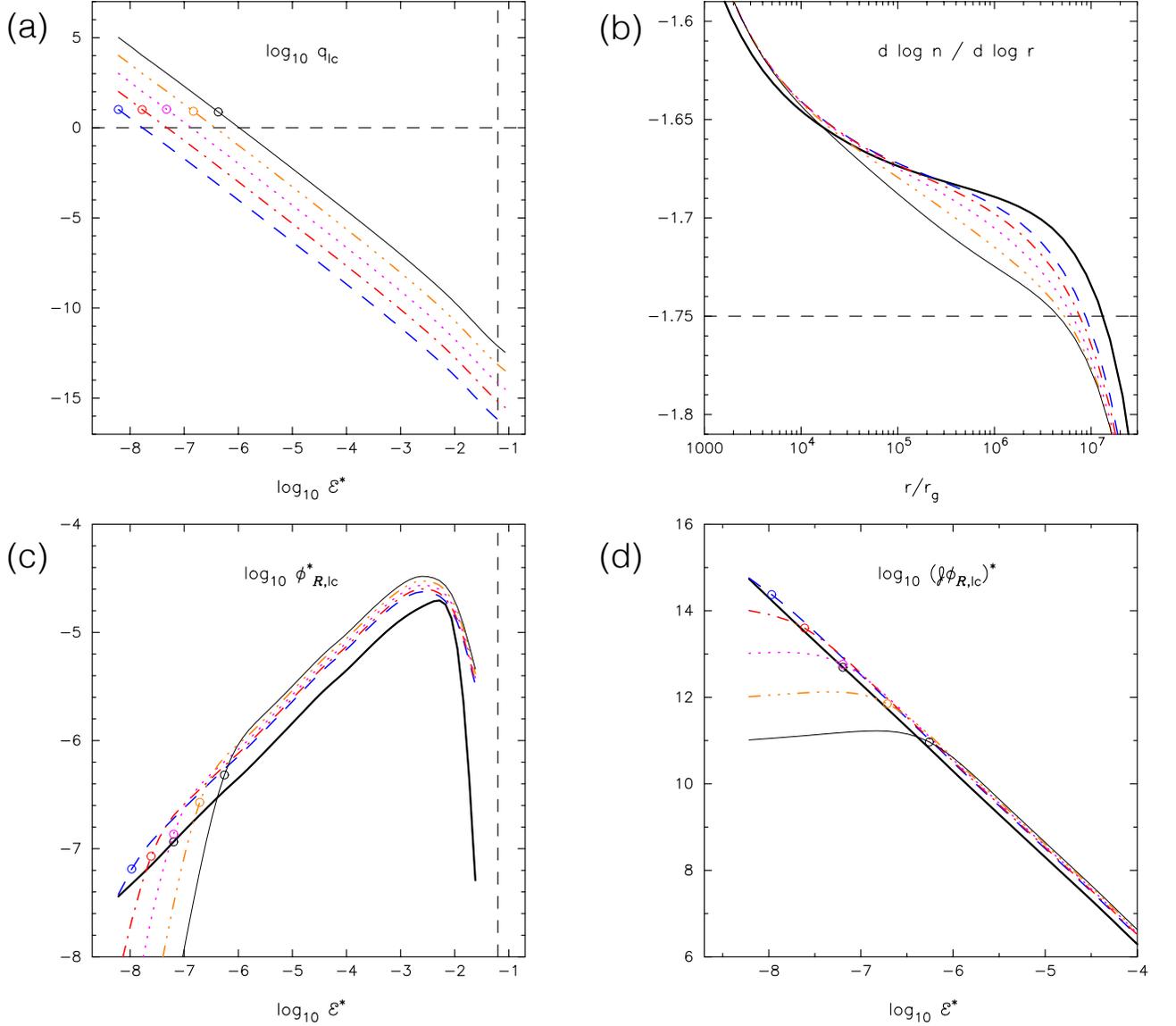}
  \caption{Final quantities for a set of integrations adopting the Cohn-Kulsrud (CK) boundary conditions,
  classical diffusion coefficients,
  and various values of $m_\star/\mh$. 
  Thick solid (black): $q_\mathrm{lc} = 0$ (empty loss cone; (b)-(d) only);
  dashed (blue): $m_\star/\mh = 10^{-7}$;
  dot-dashed (red): $10^{-6}$;
  dotted (purple): $10^{-5}$;
  dash-dot-dot (orange): $10^{-4}$;
  thin solid (black): $10^{-3}$. 
  Open circles mark the energy where $q_\mathrm{lc}=|\ln{\cal R}_\mathrm{lc}|$, 
  roughly the expected energy of transition between the empty- and full-loss-cone regimes.}
\label{Figure:Four}
\end{center}
\end{figure}
Differences in scaling will be minimized if the initial model is close to the final model.
In this section, initial conditions are chosen to be close to the Bahcall-Wolf form, 
$f\sim {\cal E}^{1/4}$.
Differences in the final state will be due to differences in the value of (\ref{Equation:qterm}),
i.e. to differing degrees of loss-cone ``fullness''.
Parameters common to all the integrations in this section were
\beq\label{Equation:NRparams}
\gamma = 7/4,\ \ \ \ \frac{r_m}{r_g}=10^9,\ \ \ \ \ln\Lambda=15 .
\eeq
Various values were adopted for of $m_\star/\mh$, implying different forms for $q_\mathrm{lc}({\cal E},t=0)$.
The initial density profile is approximately
\begin{eqnarray}\label{Equation:InitialDensity}
\rho(r) &\approx& \frac{3-\gamma}{2\pi} \frac{\mh}{r_m^3}\left(\frac{r}{r_m}\right)^{-\gamma}
\nonumber \\
&\approx& 2.0\times 10^2 \left(\frac{\mh}{10^6\msun}\right) \left(\frac{r_m}{10\;\mathrm{pc}}\right)^{-3}
\left(\frac{r}{r_m}\right)^{-7/4} \msun \mathrm{pc}^{-3}
\end{eqnarray}
and the unit of time becomes 
\beq
[t] \approx 3.0\times 10^7 \left(\frac{\mh/m_\star}{10^5}\right) \mathrm{yr} .
\eeq
These initial models are expected to reach steady states characterized by
\beq
f({\cal R}) \sim \frac{f(1)}{\ln(1/{\cal R}_\mathrm{lc})}\ln\left(\frac{\cal R}{{\cal R}_\mathrm{lc}}\right)
\eeq
for $q_\mathrm{lc} \ll 1$ (``empty loss cone'')
and
\beq
f({\cal R}) \sim \mathrm{const.}
\eeq
for $q_\mathrm{lc}\gg 1$ (``full loss cone'').

Figure \ref{Figure:Four} plots four dimensionless quantities associated with the steady-state solutions.

\smallskip
\noindent (a) $q_\mathrm{lc}({\cal E})$. The transition from empty- ($q_\mathrm{lc} \ll 1$) to
full- ($q_\mathrm{lc}\gg 1$) loss-cone regimes takes place at progressively larger binding energies, 
i.e. smaller radii, as $m_\star$ is increased.
For the smallest value of $m_\star$ adopted here ($m_\star/\mh=10^{-7}$) the final model
is essentially in the empty-loss-cone regime everywhere.

\smallskip
\noindent (b) The logarithmic derivative of the configuration-space density,  $d\log n/d\log r$.
At intermediate radii, $10^4\lesssim r/r_g\lesssim 10^7$, the slope is close to $-1.7$
in all final models, slightly shallower than in the (isotropic, scale-free) Bahcall-Wolf solution.
This is due to the progressive depletion of the loss cone near the \sbh, as in the
ELC examples above.
That depletion becomes  less severe as $m_\star$ and hence 
$q_\mathrm{lc}$ are increased and the slope
in the models with the largest $m_\star$ approaches most closely to $-7/4$.

\smallskip
\noindent (c) The magnitude of the ${\cal R}$-directed flux, 
$\phi_{\cal R} = -D_{\cal R\cal R} (\partial f/\partial {\cal R}) - D_{\cal R} f$,
evaluated at ${\cal R}={\cal R}_\mathrm{lc}({\cal E})$.
The empty-loss-cone model exhibits an approximate power-law dependence,
$\phi_{{\cal R},\mathrm{lc}}\sim {{\cal E}^*}^{1/2}$, at small ${\cal E}^*$ (large radius).
In the models with finite $m_\star$, the flux drops sharply beyond the energy
where $q_\mathrm{lc}({\cal E})\sim\ln {\cal R}_\mathrm{lc}$.

\smallskip
\noindent (d) The quantity ${\cal J}({\cal E}) \phi_{{\cal R}, \mathrm{lc}}({\cal E})$, 
whose integral $d{\cal E}$ is proportional to the integrated loss rate.
Since ${\cal J}\propto {\cal E}^{-5/2}$, ${\cal J} \phi_{{\cal R}, \mathrm{lc}} \sim {\cal E}^{-2}$
for small ${\cal E}$ in the empty-loss-cone model, implying an integrated loss rate
that diverges as ${\cal E}^{-1}\sim r$.
When $m_\star$ is non-zero,  ${\cal J} \phi_{{\cal R}, \mathrm{lc}}$ instead
``levels out'' roughly where $q_\mathrm{lc}\sim |\ln{\cal R}_\mathrm{lc}|$, implying
a finite total loss rate.

\begin{figure}[h!]
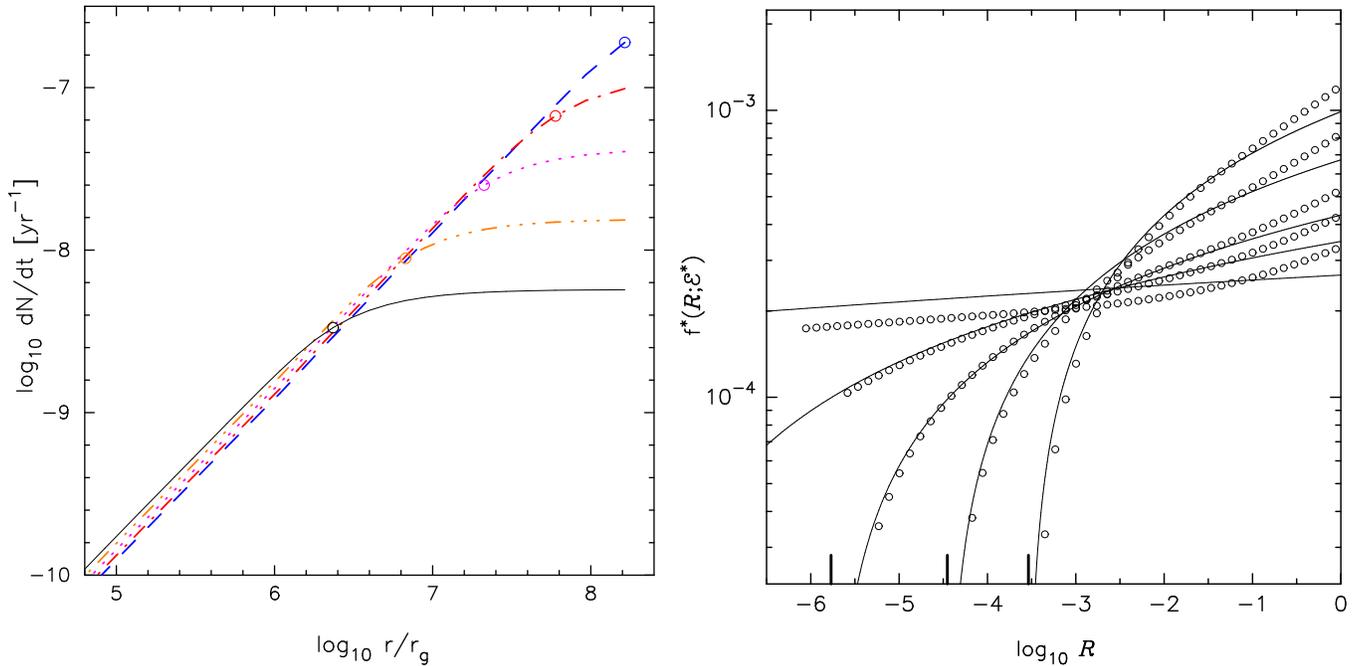

\centering
\mbox{\subfigure{\includegraphics[width=3.40in,angle=0.]{Figure8A.eps}}\quad
\subfigure{\includegraphics[width=3.5in,angle=0.]{Figure8B.eps}}}
\caption{{\bf Left panel.} 
Dimensional loss rate, defined as number of stars per year scattered into the
  \sbh\ from radii smaller than $r$,  for the five models of Figure \ref{Figure:Four}.
  ``Radius'' here means $G\mh/(2{\cal E})$ and $\mh=10^6\msun$ has been assumed.
  Line colors and types have the same meanings as in Figure \ref{Figure:Four}.
  Circles indicate the radii where $q_\mathrm{lc} = |\ln{\cal R}_\mathrm{lc}|$.
  {\bf Right panel.} ${\cal R}$-dependence of $f$ in the steady-state model from Figure \ref{Figure:Four} 
  with $m_\star/\mh=10^{-5}$ (NR diffusion coefficients). 
  Circles are $f({\cal R}; {\cal E})$ at five values of
  ${\cal E}$, for which 
  $q_\mathrm{lc}({\cal E}) \approx \{45.4, 5.44, 0.816, 6.89\times 10^{-3}, 6.14\times 10^{-5}\}$. 
  Curves are equations (\ref{Equation:fofRanalytic}), after normalizing such that the number of 
  stars at each ${\cal E}$ is the same in the numerical and analytic models. The vertical
  tick-marks at three values of ${\cal R}$ show ${\cal R}_0({\cal E})$, equation  (\ref{Equation:DefineR0}).
\label{Figure:dndt}}
\end{figure}

Equation (\ref{Equation:DefineNdot}), together with the parameters (\ref{Equation:NRparams}),
yields a relation between the dimensional and dimensionless loss rates:
\begin{eqnarray}
\frac{dN(<r)}{dt} = 4\pi\ln\Lambda \frac{c}{r_g} \left(\frac{r_m}{r_g}\right)^{2(\gamma-3)} 
\frac{dN^*(<r^*)}{dt^*}  \ \ \ \ 
\approx 3.8\times 10^{-14} \left(\frac{\mh}{10^6\msun}\right)^{-1} \frac{dN^*}{dt^*}\; \mathrm{yr}^{-1} .
\end{eqnarray}
Figure \ref{Figure:dndt}a plots this function, at the final time, for the five models with non-zero $q_\mathrm{lc}$, assuming $\mh=10^6\msun$.
We can compare these loss rates to those predicted by a simple model based on the 
assumption of an empty loss cone \cite[][equation 6.91]{DEGN}:
\begin{subequations}\label{Equation:dNdtapprox}
\begin{eqnarray}
\dot N(<r) &\approx& \frac{4\pi}{\ln {\cal R}_\mathrm{lc}} \int^r \frac{n(r)}{T_r(r)}\; r^2\; dr 
\label{Equation:dNdtapproxa}\\
&\approx& \frac{3}{4\pi} \frac{(1+\gamma)^{3/2}(3-\gamma)^2}{(9/2-2\gamma)}
\frac{N(<r_m)^2 G^{1/2} m_\star^2 \ln\Lambda}{\mh^{3/2} r_m^{3/2} \ln {\cal R}_\mathrm{lc}}
\left(\frac{r}{r_m}\right)^{9/2-2\gamma} 
\label{Equation:dNdtapproxb} \\
&\approx& 3.4\; \frac{\ln\Lambda}{\ln {\cal R}_\mathrm{lc}}\; \sqrt{\frac{G\mh}{r_m^3}}\; \frac{r}{r_m}
\label{Equation:dNdtapproxc} \\
&\approx& 5.6\times 10^{-7} \left(\frac{\mh}{10^6\msun}\right)^{-1} \frac{r}{r_m}\; \mathrm{yr}^{-1} .
\label{Equation:dNdtapproxd}
\end{eqnarray}
\end{subequations}
(The second line follows from equations (\ref{Equation:DefineTr}) and (\ref{Equation:InitialDensity});
the third line sets $\gamma=7/4$; and the final line uses the parameters (\ref{Equation:NRparams})
and approximates ${\cal R}_\mathrm{lc}$ by $r_\mathrm{lc}/r_m = 8r_g/r_m$.)
Recalling that $r_m/r_g=10^9$ for the models of Figure \ref{Figure:dndt}, we see that
equation (\ref{Equation:dNdtapprox}) correctly predicts the loss rates in the Fokker-Planck models
for the case of small $m_\star/\mh$, i.e. in the $q_\mathrm{lc}\rightarrow 0$ limit.
In the case of an empty loss cone, there is a (linear) divergence of the integrated loss rate with $r$;
when $m_\star$ is finite, there  is a transition to the $q_\mathrm{lc} > 1$ regime in which
the loss rate is given approximately by
\begin{eqnarray}
\dot N(<r) &\approx& 4\pi  \int_{r_\mathrm{crit}}^r \frac{n(r)}{P(r)}\;{\cal R}_\mathrm{lc}(r)\; r^2\; dr 
\label{Equation:dNdtFLC}
\end{eqnarray}
where $r_\mathrm{crit}$ is the smallest radius for which $q_\mathrm{lc}$ exceeds one
\cite[][equation 6.91]{DEGN}.
For these models, equation (\ref{Equation:dNdtFLC}) predicts a contribution from the full-loss-cone
regimes that scales with radius as 
\beq
\dot N(<r) \propto r_\mathrm{crit}^{-5/4} - r^{-5/4}, \ \ \ \ r > r_\mathrm{crit}
\eeq
consistent with the ``leveling out'' of the curves in Figure \ref{Figure:dndt} at radii
greater than $\sim r_\mathrm{crit}$.

It is interesting to compare the ${\cal R}$-dependence of $f$ in the final models with the
form that is often assumed when computing loss rates \citep[e.g.][]{MagorrianTremaine1998}:
\begin{subequations}\label{Equation:fofRanalytic}
\begin{eqnarray}\label{Equation:fofRanalytica}
f({\cal R}) &=& f({\cal R}_\mathrm{lc}) + \frac{f(1) - f({\cal R}_\mathrm{lc})}{\ln(1/{\cal R}_\mathrm{lc})}  \ln\left({\cal R}/{\cal R}_\mathrm{lc}\right) , \ \ \ \ 
{\cal R}_\mathrm{lc} \le {\cal R} \le 1 ,\\ \label{Equation:fofRanalyticb}
f({\cal R}_\mathrm{lc}) &=& 
\frac{f(1)}{1+q_\mathrm{lc}^{-1} \xi(q) \ln(1/{\cal R}_\mathrm{lc})}
\end{eqnarray}
\end{subequations}
Equation (\ref{Equation:fofRanalytica}) is a steady-state (constant flux) solution of the
diffusion equation in ${\cal R}$ if the low-${\cal R}$ forms of the diffusion coefficients are used,
and equation (\ref{Equation:fofRanalyticb}) follows from the Cohn-Kulsrud treatment
\citep[as given in][equations (6.59), (6.61)]{DEGN}.
Figure \ref{Figure:dndt}b plots $f({\cal R})$, at various ${\cal E}$, in the final model from
the integration with $m_\star/\mh=10^{-5}$.
Equations (\ref{Equation:fofRanalytic}) are overplotted, after normalizing to give the same total
number at each ${\cal E}$.
The agreement is reasonably good, verifying that the  boundary conditions have
been correctly implemented.
But the dependence of $f$ on ${\cal R}$ near ${\cal R}=1$ deviates from the simple
logarithmic form of equation (\ref{Equation:fofRanalytica}), due to the fact that the correct -- not
small-${\cal R}$ asymptotic -- forms of the diffusion coefficients are used in the numerical code.

In the case of a non-empty loss cone, the $f=0$ intercept of the ``external'' 
(${\cal R}>{\cal R}_\mathrm{lc}$) solution occurs at ${\cal R}={\cal R}_0<{\cal R}_\mathrm{lc}$, where
\beq\label{Equation:DefineR0}
{\cal R}_0(q_\mathrm{lc})  = {\cal R}_\mathrm{lc}({\cal E}) e^{-q_\mathrm{lc}/\xi(q_\mathrm{lc})}
\eeq
\citep[][equation 6.65]{DEGN}.
In Figure \ref{Figure:dndt}b, ${\cal R}_0$ as given by equation (\ref{Equation:DefineR0})
is indicated by vertical marks, and appears quite consistent with the value of ${\cal R}$ at 
which the numerical solutions are tending to zero.

\subsection{Models with classical and resonant diffusion coefficients}
The form of steady-state solutions derived in the preceding two sections depended on the 
value assumed for the stellar mass $m_\star$, insofar as $m_\star$ determines $q_\mathrm{lc}$.
But the dependence was found to be weak, and all of the steady-state solutions 
could be rescaled in $f$  and ${\cal E}$ to approximately the same $\overline{f}({\cal E})$.

This scale-free property is lost if the resonant relaxation diffusion terms are included.
 Sufficiently near the SBH, changes in angular momentum will be dominated by resonant relaxation
while changes in energy will still be dominated by the classical diffusion coefficients.
A characteristic radius appears: the distance, $r_\mathrm{eq}$, 
from the \sbh\ at which the timescale for changes
in $L$ due to resonant relaxation is the same as the timescale for changes in $L$ due to 
classical relaxation. 
That radius is given roughly by the solution to
\begin{equation}\label{Equation:req}
m_\star N(<r_\mathrm{eq}) = A_\mathrm{NR}\mh, \ \ \ \ 
A_\mathrm{NR} = \frac{0.68}{(3-\gamma)(1+\gamma)^{3/2}} \frac{1}{\log\Lambda}
\end{equation}
\citep[][Equation 5.243]{DEGN}.
Here $N(<r)$ is the number of stars within a sphere of radius $r$
and $t_\mathrm{coh} = t_\mathrm{coh,M}$ has been assumed;
the expression for $A_\mathrm{NR}$ additionally assumes $n(r)\propto r^{-\gamma}$.
Setting $\log\Lambda = 15$, the enclosed (stellar) mass at $r_\mathrm{eq}$ works out to be
$\sim 10^{-2} \mh$ for all $0.5\le\gamma\le 2.0$. 

Consider a star whose energy diffuses below $E_\mathrm{eq}\sim -G\mh/r_\mathrm{eq}$.
The timescale for diffusion in $L$ will suddenly decrease, and the star will be lost to the \sbh,
in a time much less than the characteristic time for changes in $E$.
A steady state will eventually be reached, but only after $|\partial f/\partial E|$ 
at $E\sim E_\mathrm{eq}$
becomes large enough to drive a flux in $E$ (due to classical relaxation) 
that equals the flux in $L$ (due to resonant relaxation).
The expected result is a depletion in $f$ at $E\sim E_\mathrm{eq}$ and a 
low configuration-space density at $r\lesssim r_\mathrm{eq}$ -- a ``core''.
\begin{figure}[h!]
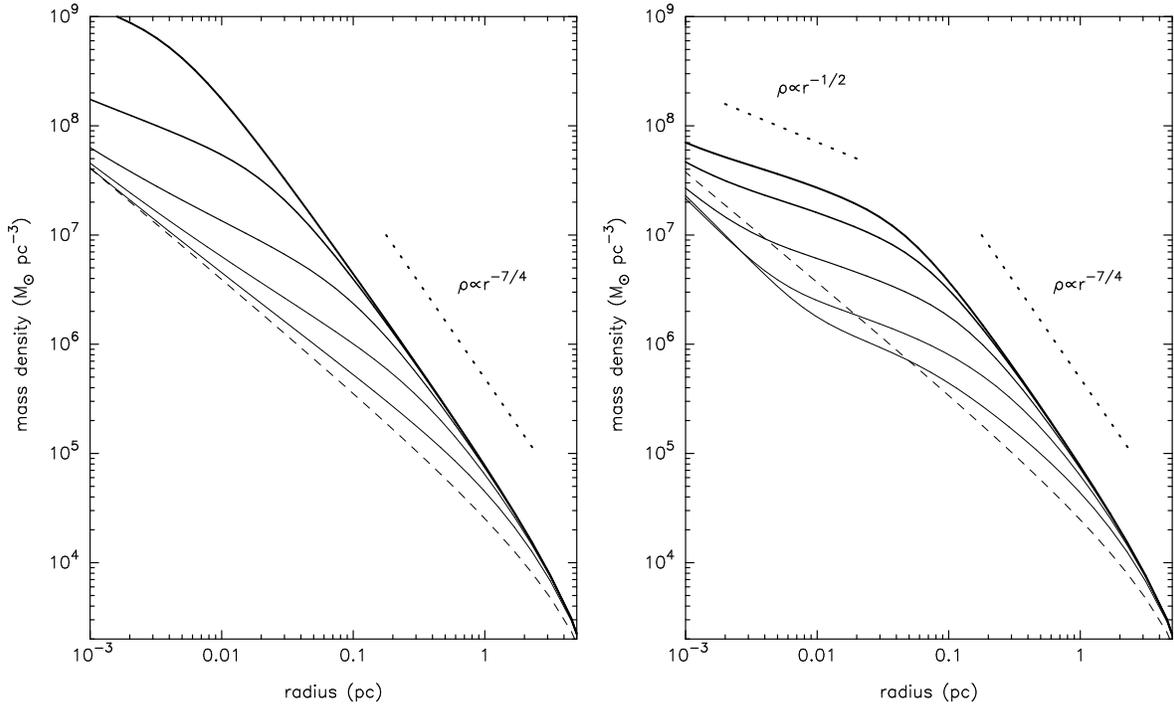

\centering
\mbox{\subfigure{\includegraphics[width=3.0in,angle=0.]{Figure9A.eps}}\quad
\subfigure{\includegraphics[width=3.0in,angle=0.]{Figure9B.eps}}}
\caption{Evolution of $\rho(r)$ in two models with the same initial density, 
$\rho(r) \propto r^{-1}$.
{\bf Left:} classical diffusion coefficients were used.
{\bf Right:} resonant relaxation terms were included as well.
Other features of the integrations are specified in the text; the adopted
parameters are appropriate for the nuclear cluster of the Milky Way, 
with $\mh=4.0 \times 10^6\msun$.
Densities are plotted at times $t = \{0,2,4,6,8,10\}\times 10^9$ yr.
The $t=0$ model is shown by dashed lines, and line width increases with time.
Dotted lines show the Bahcall-Wolf slope, $d\log\rho/d\log r=-7/4$,
and the slope corresponding to a sharply truncated $f({\cal E})$, 
$d\log\rho/d\log r = -1/2$.
Both models have an integrated stellar mass, within one parsec, of approximately
$1.0\times 10^6\msun$ at $t=5\times 10^9$ yr, consistent with the mass measured in the
Milky Way nuclear cluster \citep{Schoedel2009}.
\label{Figure:tworho}}
\end{figure}

Figure \ref{Figure:tworho} shows the evolution of $\rho(r)$ in two integrations:
the first using the classical diffusion coefficients, 
the second including the resonant-relaxation coefficients as well.
Both integrations adopted parameters appropriate for the nuclear cluster of the Milky Way.
The quantity $(m_\star/\mh)\ln\Lambda$ that appears in equation (\ref{Equation:qlcofFstar}) 
for $q_\mathrm{lc}$:
\beq
\frac{m_\star}{\mh}\ln\Lambda = 3.75\times 10^{-6} \left(\frac{m_\star}{1\msun}\right)
\left(\frac{\mh}{4\times 10^6\msun}\right)^{-1}
\left(\frac{\ln\Lambda}{15}\right)
\eeq
was set to $3.75\times 10^{-6}$, and the radius of the loss sphere, which for a
main-sequence star is the tidal disruption radius:
\begin{equation}\label{Equation:DefinerLC}
r_\mathrm{lc}\approx 0.7{\rm AU}~ \frac{R_\star}{R_\odot} \left( {M_{\bullet} \over m_{\star}}{1 \over 4\times10^6}  \right)^{1/3}
\end{equation}
was set to $0.7$ AU.
These choices were motivated by the fact that the main-sequence turnoff mass in the
Galactic center is $\sim 1\msun$, and the red giants that are believed to dominate
the number counts of the ``late-type'' stars probably have roughly this mass
 \citep{Dale2009}.
(Due to the relative shortness of the red giant evolutionary phase, most stars
are expected to be disrupted while still on or near the main sequence
\citep{MacLeod2012}).
These choices determined the location and flux at the loss-cone boundary via
the expressions (\ref{Equation:FluxLCR}), (\ref{Equation:FluxLCE}).

The initial density normalization was chosen, after some trial and error, to give
an integrated mass within $1$ pc of $\sim 1.0\times 10^6\msun$ at $t=5\times 10^9$ yr.
This is roughly the value inferred from dynamical analyses of stellar velocities 
\citep{Schoedel2009}.

Figure~\ref{Figure:tworho} shows that the integration based on the classical diffusion coefficients
nearly reaches the $\rho\propto r^{-7/4}$ Bahcall-Wolf form after 10 Gyr.
Deviations from that form are apparent inside $\sim 0.01$ pc even at this late time; 
however seen from the Earth, that radius would subtend an angle of only $\sim 0.^{\prime\prime}25$.

Inclusion of the resonant-relaxation diffusion coefficients implies a different steady state.
Inside $\sim 0.05$ pc, the density profile remains much shallower than the Bahcall-Wolf form,
with $\rho \sim r^{-0.5}$ -- the functional form that corresponds to an $f$ that is fully depleted
at high binding energies.
The radius of this ``core'' is consistent with the prediction made above ($\sim 0.01 r_m$)
given that $r_m\approx 2.5$ pc.

Since the radius of the core is a function of the density normalization, 
steady state solutions in the presence of resonant relaxation are expected to depend on (at least)
one more parameter than in the classical case.
A more general exploration of solutions like these will be presented in Paper II from this series
 \citep{PaperII}.


\acknowledgements
I thank A. Hamers for providing data from his {\tt TPI} code that was used in deriving
the functional forms of the RR diffusion coefficients in \S \ref{Section:RR}, and H. Cohn for
helpful discussions.
This work was supported by the National Science Foundation under grant no. AST 1211602 
and by the National Aeronautics and Space Administration under grant no. NNX13AG92G.


 \appendix
  \renewcommand{\theequation}{A\arabic{equation}}
  \setcounter{equation}{0} 
  \centerline{Numerical Evaluation of the $C_i$} 
   \label{Appendix:Ci}
\bigskip

The functions $C_i(s={\cal E}^\prime/{\cal E}, {\cal R})$ 
that appear in equation (\ref{Equation:Fii}),
\begin{equation}
F_i\left({\cal E}, {\cal R}\right) = 4\pi\Gamma\int_{\cal E}^{{\cal E}/x_-} 
\overline{f}\left({\cal E}^\prime\right) C_i\left(\frac{{\cal E}^\prime}{\cal E}, {\cal R}\right)
d{\cal E}^\prime \nonumber
\end{equation}
are given by \citet{CohnKulsrud1978} as
\begin{eqnarray}
C_1 &=& \frac{2}{\pi} \int dx\; Q^{-1/2}\; \frac{x(1-sx)^{1/2}}{(1-x)^{1/2}} ,\nonumber \\
C_2 &=& \frac{2}{\pi} \int dx\; Q^{-1/2}\; \frac{x^2(1-sx)^{1/2}}{(1-x)^{3/2}} ,\nonumber \\
C_3 &=& \frac{2}{\pi} \int dx\; Q^{-1/2}\; \frac{x^3(1-sx)^{1/2}}{(1-x)^{1/2}} ,\nonumber \\
C_4 &=& \frac{2}{\pi} \int dx\; Q^{-1/2}\; \frac{(1-sx)^{3/2}}{(1-x)^{1/2}} ,\nonumber \\
C_5 &=& \frac{2}{\pi} \int dx\; Q^{-1/2}\; \frac{x(1-sx)^{3/2}}{(1-x)^{3/2}} ,\nonumber \\
C_6 &=& \frac{2}{\pi} \int dx\; Q^{-1/2}\; \frac{x^2(1-sx)^{3/2}}{(1-x)^{5/2}} ,\nonumber \\
C_7 &=& \frac{2}{\pi} \int dx\; Q^{-1/2}\; \frac{x^3(1-sx)^{3/2}}{(1-x)^{3/2}} .
\end{eqnarray}
In these expressions, $Q\equiv (x_+-x)(x-x_-)$ with 
$x_{\pm}=(1/2)\left[1\pm\sqrt{1-{\cal R}}\right]$, and the limits of integration are
$x_-$ and $\max\left[x_-,\min\left(x_+,s^{-1}\right)\right]$.
These functions are independent of $f$ and so can be evaluated on a fixed numerical grid.
The grid axes were chosen to be $\{{\cal R}, W\}$ where
\begin{eqnarray}
W \equiv \frac{2T - 1 + \sqrt{1-{\cal R}}}{1+\sqrt{1-{\cal R}}}
\end{eqnarray}
and $T \equiv s^{-1} = {\cal E}/{\cal E}^\prime$.
Thus $0\le W\le 1$, $0\le {\cal R}\le 1$, and
\begin{equation}
T = \frac12\left[1-\sqrt{1-{\cal R}} + W\left(1+\sqrt{1-{\cal R}}\right)\right] .
\end{equation}
Typically the number of grid points was $256\times 256$.

The $C_i$ are finite for all $\{{\cal R}, W\}$ with the exception of $C_2$ and $C_6$, 
which diverge at 
\begin{equation}
{\cal R} = 0, \ \ \ \ W=T=s=1.
\end{equation}
Along the ${\cal R}=0$ border, both functions diverge as $\left(1-W\right)^{-1/2}$ as
$W\rightarrow 1$, while along the $W=1$ border they diverge as ${\cal R}^{-1/2}$ as
${\cal R}\rightarrow 0$.
To deal with the divergence, $C_2$ and $C_6$ were multiplied by the function
\begin{equation}
\left[{\cal R}^2 + 256\left(1-W\right)^2\right]^{1/4}
\end{equation}
before storing their computed values on the grid.
Integrations were carried out with routine {\tt d01apf} from the
{\it Numerical Algorithms Group} fortran subroutine library.
Accuracy of the integrations was checked by comparison with analytic expressions
that obtain along the grid boundaries; for instance, for ${\cal R}=1$,
$C_i=g_i(s)$ where
\begin{eqnarray}
g_1 &=& g_2 = \sqrt{2-s}, \ \ \ \ 
g_3 = \frac14\sqrt{2-s},\nonumber \\
g_4 &=& g_5 = g_6 = \left(2-s\right)^{3/2}, \ \ \ \ 
g_7 = \frac14\left(2-s \right)^{3/2} .
\end{eqnarray}

Once the values of the $C_i$ had computed on the $\{{\cal R},W\}$ grid,
the  NAG routine {\tt e01daf} was used to fit a bicubic interpolating spline to the computed values.
 During integrations of the Fokker-Planck equation,
values of the $C_i$ between the $\{{\cal R}, W\}$ grid points
were then computed from the spline coefficients.

  \appendix
  \renewcommand{\theequation}{B\arabic{equation}}
  \setcounter{equation}{0} 
  \centerline{$N_r(a)$ vs. $N_a(a)$} 
  \label{Appendix:NvsN}
  \bigskip
  This appendix compares two quantities:
  \medskip
  \begin{enumerate}
 \item $N_r(<a)$, the number of stars with instantaneous radii less than $a$;
 \item $N_a(<a)$, the number of stars with semimajor axes less than $a$.
\end{enumerate}  

A  power-law dependence of number density on distance from the \sbh\ is assumed:
\beq
n(r) =n_0\left(\frac{ r}{r_0}\right)^{-\gamma}, \ \ \ \ \psi(r) = \frac{G\mh}{r} 
\eeq
so that the number of stars instantaneously below $r$ is
\beq
N_r(<r) = \frac{4\pi}{3-\gamma} n_0r_0^3 \left(\frac{r}{r_0}\right)^{3-\gamma} .
\label{Equation:Nofr}
\eeq
The distribution function is assumed to be isotropic; Eddington's formula gives
\begin{subequations}
\begin{eqnarray}
f({\cal E}) &=& f_0\left(\frac{{\cal E}}{{\cal E}_0}\right)^\eta,\ \ \ \ \eta = \gamma - 3/2 \\
&=& \frac{1}{\left(2\pi\right)^{3/2}} \frac{\Gamma(\gamma+1)}{\Gamma(\gamma-1/2)}
\frac{n_0r_0^\gamma}{\left(G\mh\right)^{-\gamma}} \;{\cal E}^{\gamma-3/2} .
\end{eqnarray}
\end{subequations}
The number of stars per unit of binding energy is
\begin{eqnarray}
N({\cal E}) d{\cal E} &=& 4\pi^2 p({\cal E}) f({\cal E})\; d{\cal E},\ \ \ \ 
p({\cal E}) = \frac{\sqrt{2}\pi}{4} \left(G\mh\right)^3 {\cal E}^{-5/2}
\end{eqnarray}
so that
\beq
N({\cal E}) d{\cal E} = \frac{\pi^{3/2}}{2} \frac{\Gamma(\gamma+1)}{\Gamma(\gamma-1/2)}
\left(G\mh\right)^{3-\gamma} n_0r_0^{\gamma}\; {\cal E}^{\gamma-4} d{\cal E}
\eeq
and the number of stars with binding energies greater than ${\cal E}$ is
\begin{eqnarray}
N_{\cal E}(>{\cal E}) = \int_{\cal E}^{\infty} N({\cal E}) d{\cal E} 
= \frac{\pi^{3/2}}{2(3-\gamma)} \frac{\Gamma(\gamma+1)}{\Gamma(\gamma-1/2)}
\left(G\mh\right)^{3-\gamma}n_0r_0^{\gamma}\; {\cal E}^{\gamma-3} .
\end{eqnarray}
Setting ${\cal E}=G\mh/(2a)$ yields the number of stars with semimajor axes less than $a$:
\begin{eqnarray}
N_{a}(<a) =  \frac{2^{2-\gamma}\pi^{3/2}}{(3-\gamma)} 
\frac{\Gamma(\gamma+1)}{\Gamma(\gamma-1/2)}
n_0r_0^3\; \left(\frac{a}{r_0}\right)^{3-\gamma} .
\label{Equation:Nofa}
\end{eqnarray}

Equations (\ref{Equation:Nofr}) and (\ref{Equation:Nofa}) can be written
\begin{eqnarray}
N_r(<a) = N_1(\gamma) \; n_0r_0^3 \; \left(\frac{a}{r_0}\right)^{3-\gamma}, \ \ \ \ 
N_a(<a) = N_2 (\gamma)\; n_0r_0^3 \; \left(\frac{a}{r_0}\right)^{3-\gamma}
\end{eqnarray}
with
\begin{eqnarray}
N_1(\gamma) = \frac{4\pi}{3-\gamma} , \ \ \ \ 
N_2(\gamma) &=& \frac{2^{2-\gamma} \pi^{3/2}}{3-\gamma} \frac{\Gamma(\gamma+1)}{\Gamma(\gamma-1/2)}.
\end{eqnarray}
\begin{table}[h]
\centering 
\begin{tabular}{c c c c c c} 
\hline\hline 
$\gamma$ & 1 & 3/2 & 7/4& 2 & 5/2\\ [0.5ex] 
\hline 
$N_1$ & 6.28 & 8.38 & 10.05 & 12.56 & 25.13 \\ 
$N_2$ & 3.14 & 6.98 & 9.40 & 12.56 & 26.17 \\
\hline
$\boldsymbol{N_2/N_1}$ & \bf{0.50} & \bf{0.83} & \bf{0.94} & \bf{1.00} & \bf{1.04}  \\
\hline 
\end{tabular}
\end{table}
Values of $N_1$ and $N_2$ are given in the table.
For $\gamma \gtrsim 3/2$ the two quantities are very similar; they begin to depart significantly for
smaller $\gamma$.

  \appendix
  \renewcommand{\theequation}{C\arabic{equation}}
  \setcounter{equation}{0} 
  \centerline{Anomalous Relaxation} 
  \label{Appendix:AR}
  \bigskip
A Monte-Carlo algorithm for describing the evolution of $L$ in the ``anomalous relaxation''
($L\lesssim L_\mathrm{SB}$) regime was presented in \citet{MAMW2011}.
That algorithm was based on a simple model for the time- and space-dependence
of the $\sqrt{N}$ perturbing potential.
Following is an analytic derivation of the angular momentum transition probabilities, 
and the corresponding diffusion coefficients, that are implied by the same Hamiltonian model.
The results of the derivation presented in this appendix were the basis for the functional forms that \citet{Hamers2014} fit to diffusion coefficients extracted from their $N$-body data, and which appear here in \S \ref{Section:AR}.

Consider a star orbiting in the potential
\beq
\Phi(\boldsymbol{r}) = \Phi_\mathrm{Kepler} + \Phi_\mathrm{N} + \Phi_\mathrm{\sqrt{N}} \;.
\eeq
Here, $\Phi_\mathrm{Kepler} = -G\mh/r$ is the (Newtonian) potential due to the \sbh;
$\Phi_\mathrm{N}$ is the potential from the spherically-distributed mass;
and $\Phi_\mathrm{\sqrt{N}}$ is the potential due to the $\sqrt{N}$ asymmetries in the
stellar distribution.
We assume that $|\Phi_{\sqrt{N}}| \ll |\Phi_\mathrm{N}| \ll |\Phi_\mathrm{Kepler}|$.

If the mass density falls off as a power of radius, $\rho(r)=\rho_0(r/r_0)^{-\gamma}$, then
\beq
\Phi_\mathrm{N}(r) = \frac{4\pi G\rho_0 r_0^2}{(2-\gamma)(3-\gamma)} \left(\frac{r}{r_0}\right)^{2-\gamma}
+ \mathrm{constant}
\eeq
for $\gamma\ne 2$; for $\gamma=2$ the dependence of $\Phi_\mathrm{N}$ on radius 
becomes logarithmic.
Following \citet{MAMW2011}, we assume that the $\sqrt{N}$ perturbing potential, as
experienced by a test star of semimajor axis $a$, is given by 
\begin{eqnarray}
\Phi_\mathrm{\sqrt{N}} = -a S(a) \cos\theta = -a S(a) \frac{z}{r} ,\ \ \ \ 
S(a) = \frac{G m_\star}{a^2} \sqrt{N(a)} \;.
\end{eqnarray}
We are assuming for the moment that $\Phi_{\sqrt{N}}$ is independent of time.
Expressing the two perturbing potentials in Delaunay variables and averaging over the unperturbed
(Keplerian) motion yields
\begin{subequations}
\begin{eqnarray}
\overline\Phi_\mathrm{N} &=&\frac{GM_\star(r<a)}{(2-\gamma)a} 
\left(1 +\alpha_1 -\alpha_2\ell^2\right) ,\\
\overline\Phi_{\sqrt{N}} &=& a S(a) e \sin i \sin\omega .
\end{eqnarray}
\end{subequations}
Here, $\ell\equiv \sqrt{1-e^2}$, $i=\pi/2$ corresponds to the $x-z$ plane and 
$\omega=\pi/2$ describes an orbit that is elongated along $z$.
The expression for $\overline\Phi_\mathrm{N}$ assumes $\ell\ll 1$; the quantities
$\alpha_1(\gamma)$, $\alpha_2(\gamma)$ are both of order unity and are
given in \citet[][\S4.4.1]{DEGN}.

Ignoring constant terms (including terms that depend only on $a$),
the averaged Hamiltonian is then
\begin{subequations}
\begin{eqnarray}
H \equiv \frac{\overline\Phi}{\nu_0I} &=& -\ell^{-1} - A_\mathrm{N}\ell^2 + 
A_\mathrm{\sqrt{N}} \; e \sin i \sin\omega,\\
A_\mathrm{N} &=& \frac{\alpha_2}{3(2-\gamma)}\frac{M_\star(a)}{\mh} \frac{a}{r_g},\ \ \ \ 
A_\mathrm{\sqrt{N}} = \frac12\frac{S(a)}{G\mh/a^2}\frac{a}{r_g}
\end{eqnarray}
\end{subequations}
and
\beq
r_g\equiv \frac{G\mh}{c^2},\ \ \ \nu_0 = \frac{2\pi}{P(a)} \frac{3r_g}{a}, \ \ \ I = (G\mh a)^{1/2}.
\eeq
Discarding terms of order $\ell^2$ or smaller, the averaged Hamiltonian becomes
\beq\label{Equation:Hsimple}
H = -\ell^{-1} + A_{\sqrt{N}}\sin i \sin \omega.
\eeq
Note that the ``mass precession'' terms vanishes to this order in $\ell$.
The first term in equation (\ref{Equation:Hsimple}) represents GR precession; the
second represents the effects of the $\sqrt{N}$ torques.

Writing $\tau = \nu_0t$, Hamilton's equations of motion for the osculating elements are
\begin{eqnarray}\label{Equation:EOM}
\frac{d\omega}{d\tau} &=& \frac{\partial H}{\partial \tau} = \ell^{-2}, \ \ \ \ 
\frac{d\ell}{d\tau} = -\frac{\partial H}{\partial\omega} = -A_{\sqrt{N}}\sin i \cos\omega, \nonumber \\
\frac{d\Omega}{d\tau} &=& \frac{\partial H}{\partial \ell_z} = -A_{\sqrt{N}}
\frac{\ell_z}{\ell^2}\frac{\sin\omega}{\sin i}, \ \ \ \ 
\frac{d\ell_z}{d\tau} = 0.
\end{eqnarray}
Henceforth setting $\sin i = 1$, 
the dependence of $\ell$ on $\omega$ is given by
\begin{subequations}
\begin{eqnarray}
\ell^{-1} &=& \frac{1}{2\ell_1\ell_2}\left[(\ell_2-\ell_1)\sin\omega + (\ell_1+\ell_2)\right]\\
&=& A_{\sqrt{N}}\left(\sin\omega + g_0\right)
\end{eqnarray}
\end{subequations}
where $\{\ell_1,\ell_2\}$ are the extreme values of $\ell$ and 
$g_0=-H/A_{\sqrt{N}}$ is the ``energy.''
The equation of motion for $\ell$ can be expressed in terms of $\ell$ alone as
\beq
\frac{d\ell}{d\tau} = \pm \sqrt{\left(\frac{1}{\ell_1}-\frac{1}{\ell}\right)
\left(\frac{1}{\ell_1}-\frac{1}{\ell}\right)},
\eeq
with a similar expression for $\dot \omega$, and 
the full precessional period is
\beq
T_0 = \frac{2\pi}{A_{\sqrt{N}}^2}\frac{g_0}{\left(g_0^2-1\right)^{3/2}} 
\eeq
where 
\beq
\ell_1 = A_{\sqrt{N}}^{-1}\left(g_0+1\right)^{-1},\ \ 
\ell_2 = A_{\sqrt{N}}^{-1}\left(g_0-1\right)^{-1} ;
\eeq
thus $g_0 = (\ell_2 + \ell_1)/(\ell_2-\ell_1)$ and 
\beq\label{Equation:Defineellav}
\frac{\ell_1 + \ell_2}{2} \equiv \ell_\mathrm{av} = \frac{1}{A_{\sqrt{N}}} \frac{g_0}{g_0^2-1} .
\eeq

Solutions obtained so far describe ``coherent resonant relaxation:'' 
changes in a star's angular momentum for times shorter than the coherence time.
Now, suppose that the orientation of the torquing potential changes, instantaneously, 
at random times separated by $t_\mathrm{coh}$.
Of course this is a crude oversimplification since in reality the torquing potential is changing
gradually; on the other hand, for stars near the 
Schwarzschild barrier, the GR precession time is expected to be comparable to the coherence time
(equation~\ref{Equation:SB2}).

Let the angle between the new $z$-axis , and the orbital
semimajor axis at the moment of the switch, be $\omega_1$.
At this moment, $\ell$ has the value $\ell_s$, and $g_0$ changes to
$g_1$, where
\beq
g_1 = 1/(A_{\sqrt{N}}\ell_s) - \sin\omega_1.
\eeq
Since the probability distributions of $\tau$ and $\omega_1$ are uniform,
we can write
\beq
P\left(\ell_s,g_1\right)d\ell_sd g_1 = P\left(\tau,\omega_1\right) d\tau d\omega_1
= \frac{4}{2\pi T_0}.
\eeq
\begin{figure}
  \begin{center}
    \includegraphics[width=8cm]{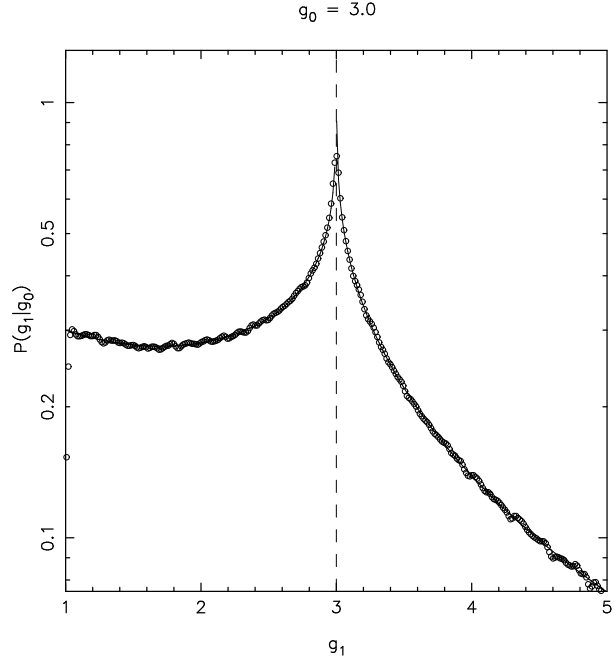}
  \caption{\label{Figure:Pg1} $P(g_1|g_0)$ computed in two ways, for $g_0=3$.
  }\end{center}
  \label{fig:pg1g0}
\end{figure}
\noindent The factor four on the RHS accounts for the fact that $\ell$ varies over its
full range in one-fourth of a precessional period.
Then
\begin{subequations}
\begin{eqnarray}
P(\ell_s,g_1) &=& \frac{2}{\pi T_0}\frac{\partial(\tau,\omega_1)}{\partial(\ell_s,g_1)}
\equiv \frac{2}{\pi T_0} J, \\
J^{-1} &=& \frac{\partial(\ell_s,g_1)}{\partial(\tau,\omega_1)} 
= \left|\frac{d\ell_s}{d\tau}\frac{\partial g_1}{\partial\omega_1}\right|, \\
J &=& A_{\sqrt{N}}\sqrt{1-\left[g_0 -1/(A_{\sqrt{N}}\ell_s)\right]^2}
\sqrt{1-\left[g_1 -1/(A_{\sqrt{N}}\ell_s)\right]^2}.
\end{eqnarray}
\end{subequations}
Integrating over $\ell_s$ gives the probability distribution for $g_1$.
This can be written
\begin{eqnarray}\label{Equation:Pint}
P(g_1|g_0) = \frac{2}{\pi T_0A_{\sqrt{N}}^2}\int_{x_1}^{x_2}
\frac{dx}{x^2\sqrt{(x-x_1)(x_2-x)\left[2-(x-x_1)\right]
\left[2-(x_2-x)\right]} }
\end{eqnarray}
where
\beq\label{Equation:Defx1x2a}
x_1 = g_0-1, \ \ \ x_2 = g_1 + 1
\eeq
for $g_1<g_0$, and
\beq\label{Equation:Defx1x2b}
x_1 = g_1-1, \ \ \ x_2 = g_0 + 1
\eeq
for $g_1>g_0$.
Figure~\ref{Figure:Pg1} shows a numerical integration of equation~(\ref{Equation:Pint})
(solid line) for $g_0=3$, compared with the results of
Monte-Carlo experiments based on the equations of motion (\ref{Equation:EOM}).
We note here the asymmetry of the derived transition probability.

The diffusion coefficients are
\beq\label{Equation:Diff}
\langle(\Delta g_0)^k\rangle = \frac{1}{T_\mathrm{coh}}
\int_{g_0-2}^{g_0+2}\left(g_1-g_0\right)^k P\left(g_1|g_0\right) dg_1.
\eeq
The integral (\ref{Equation:Diff}) can be broken into two pieces, $I_>$ and $I_<$,
corresponding to $g_0>g_1$ and $g_0<g_1$ respectively.
In the case of $\langle\Delta g_0\rangle$,
\begin{subequations}
\begin{eqnarray}
\left(\frac{\pi}{2} T_0A_{\sqrt{N}}^2T_\mathrm{coh}\right) I_> &=&
\int_{x_1}^{x_1+2}\left(x_2-x_1-2\right) dx_2 
\int_{x_1}^{x_2}\frac{dx}{x^2}
\frac{1}{\sqrt{(x_2-x)(x-x_1)\left[2-(x_2-x)\right]
\left[2-(x-x_1)\right]}} 
\\
&=& -\frac{2}{g_0^2-1} - 2\int_0^2\frac{(1-w)dw}{\sqrt{w(2-w)}\left(w+g_0-1\right)^2}
\sin^{-1}\sqrt{\frac{2-w}{2}}
\end{eqnarray}
\end{subequations}
and a similar calculation gives
\begin{equation}
\left(\frac{\pi}{2} T_0A_{\sqrt{N}}^2T_\mathrm{coh}\right) I_< =
 \frac{2}{g_0^2-1} + 2\int_0^2\frac{(1-w)dw}{\sqrt{w(2-w)}\left(-w+g_0+1\right)^2}
\sin^{-1}\sqrt{\frac{2-w}{2}}.
\end{equation}
The sum is
\begin{subequations}\label{Equation:Ig0}
\begin{eqnarray}
\left(\frac{\pi}{2} T_0A_{\sqrt{N}}^2T_\mathrm{coh}\right)\left(I_<+I_>\right) 
&=& -\frac{8}{g_0^3} I(g_0),\\
I(g_0) &=& \int_{-1}^1\frac{x^2\, dx}{(g_0^2-x^2)^2\sqrt{1-x^2}}
\sin^{-1}\sqrt{\frac{1-x}{2}} 
= \frac{\pi^2}{8} \frac{g_0^3}{\left(g_0^2-1\right)^{3/2}} .
\end{eqnarray}
\end{subequations}
Thus
\begin{equation}
\langle\Delta g_0\rangle = -\frac{8}{g_0^3} \frac{2}{\pi T_\mathrm{coh} T_0 A_{\sqrt{N}}^2} 
I(g_0)\\
= -\frac{1}{t_\mathrm{coh}}\frac{1}{g_0}.
\end{equation}

Proceeding as before to evaluate $\langle(\Delta g_0)^2\rangle$:
\begin{eqnarray}
\langle(\Delta g_0)^2\rangle &=& \frac{2}{\pi T_0 A_{\sqrt{N}}^2T_\mathrm{coh}}\left(I_<+I_>\right)
= \frac{1}{T_\mathrm{coh}} K(g_0), \nonumber \\
K(g_0) &=& \left[\frac52+\frac{\left(g_0^2-1\right)^{3/2}}{g_0} - g_0^2\right].
\end{eqnarray}
The function $K(g_0)$  is almost independent of $g_0$:
\beq
K(3) = 1.042, \ \ K(5) = 1.015, \ \ K(8) =1.006 , \ \ K(10) = 1.004, \ \ K(20) = 1.001,\ \ 
K(\infty) = 1 \nonumber
\eeq
so that
\begin{eqnarray}
\langle(\Delta g_0)^2\rangle = (1.0^{+0.042}_{-0}) \frac{1}{t_\mathrm{coh}}\; .
\end{eqnarray}
To a good approximation then,
\begin{equation}
\langle\Delta g_0\rangle = -\frac{1}{t_\mathrm{coh}}\frac{1}{g_0},\ \ \ \ 
\langle(\Delta g_0)^2\rangle = \frac{1}{t_\mathrm{coh}}
\end{equation}
and the time scales for changes in $g_0$ are 
\beq
\left|\frac{\langle\Delta g_0\rangle}{g_0}\right|^{-1} \approx 
\left|\frac{\langle(\Delta g_0)^2\rangle}{g_0^2}\right|^{-1} \approx g_0^2 t_\mathrm{coh} .
\eeq

Identifying $\ell$ with $\ell_\mathrm{av}$ (equation~\ref{Equation:Defineellav}),
diffusion coefficients in $\ell$ become
\begin{subequations}\label{Equation:DiffARellexact}
\begin{eqnarray}
\langle\Delta\ell\rangle &\approx& \frac{2\ell^3}{\tau} \left[1+\frac32\left(A_{\sqrt{N}}\;\ell\right)^2\right] , \\
\langle\left(\Delta\ell\right)^2\rangle &\approx& \frac{\ell^4}{\tau} \left[1+2\left(A_{\sqrt{N}}\;\ell\right)^2\right] 
\end{eqnarray}
\end{subequations}
where $\tau(a) \equiv t_\mathrm{coh}(a)/(A_\mathrm{\sqrt{N}})^2$.

\end{document}